\documentclass[sn-basic,numbers]{sn-jnl}
\usepackage{graphicx} 
\usepackage{subcaption}
\usepackage{amsmath,bbm}
\usepackage[textwidth=2cm]{todonotes}
\setuptodonotes{color=green!50,fancyline}
\newcommand{\RR}{\mathbbm{R}}

\begin{document}

\title{Sympatric speciation by symmetry-breaking: 

The three-clade case}

\author[1]{\fnm{Giagkos-Ion} \sur{Chlomoudis}}\email{G-I.Chlomoudis@uni-kassel.de}
\author[1,3]{\fnm{Thomas} \sur{Fuhrmann-Lieker}}\email{th.fuhrmann@uni-kassel.de}
\author[2]{\fnm{Meskerem A.} \sur{Mebratie}}\email{meskerem.mebratie@uni-kassel.de}
\author[2]{\fnm{Gokul B.} \sur{Nair}}\email{gokul.nair@uni-kassel.de}
\author[2]{\fnm{Werner M.}
  \sur{Seiler}}\email{seiler@mathematik.uni-kassel.de}

\affil[1]{\orgdiv{Institute of Chemistry}, \orgname{University of Kassel},
  \city{Kassel},  \postcode{34109}, \country{Germany}}
\affil[2]{\orgdiv{Institute of Mathematics}, \orgname{University of Kassel},
  \city{Kassel},  \postcode{34109}, \country{Germany}}
\affil[3]{\orgdiv{Center for Interdisciplinary Nanostructure Science and
    Technology}, \orgname{University of Kassel}, \city{Kassel},
  \postcode{34109}, \country{Germany}}

\abstract{In this paper we expand the concept of biological speciation by
  symmetry breaking of Golubitsky and Stewart to the case of three clades
  in which $N$ populations following the same dynamical laws can
  separate. The underlying differential equation is based on a fifth order
  polynomial of a trait variable with first or second order coupling. We
  present some general strategies to find all possible steady states and
  their stabilities. Numerical data are given for a specific system. We
  show the locations of three-clade distributions in dependence on the
  coupling and an environmental parameter. The results show a decrease of
  the number of stable states with higher coupling and a higher probability
  of ending in a three-clade state for larger $N$. Limits and potentials of
  the approach if zero roots for the trait variable occur are discussed.}



\maketitle

\section{Introduction}

Sympatric speciation, according to which new species evolve from a single
ancestral species while inhabiting the same geographic region, remains a
debated topic in evolutionary biology
\citep{kautt2020,malinsky2015,osborne2020}. It has been reported
throughout the tree of life (bacteria, fungi, plants, animals, etc.), and
in some taxa it may even be particularly common
\citep{bushsmith1998,hernandez2021}. In many animals, for instance,
particularly phytophagous insects or parasitic organisms, a shift to a new
host or habitat can trigger divergence, often requiring only a small number
of genetic changes when mating occurs within the preferred habitat or
resource \citep{bush1994}. In the lakes of East Africa, hundreds of endemic
Cichlid fish species have evolved in a relatively short evolutionary time
without strict geographic isolation
\citep{orrSmith1998,schluter1998}. Similar patterns occur in postglacial
lakes, where distinct ecological forms of fish exhibit morphological and
ecological differentiation while remaining genetically similar at neutral
markers\citep{orrSmith1998}. In plants, sympatric speciation can occur
through polyploidy, which can produce immediate reproductive isolation and
may account for approximately $4\%$ of newly formed angiosperm species
\citep{coyne2007}. For the pennate diatom \textit{Seminavis robusta}, three
distinct clades with less successful inter-clade crosses than intra-clade
crosses are found and interpreted by incomplete reproductive isolation
\citep{dedecker2018}. The observation that two of the clades are more
closely related to each other than the third one suggests two distinct
speciation events that are separated by several hundred thousand years
\citep{vyverman2024}. Such intermediates between sympatric and allopatric
speciation --- characterized by the coupling of geographic locations
including the possibility of gene transfer --- are usually called
parapatric speciation.

The existence of sym- and parapatric speciation proves that geographic
isolation is not a prerequisite for speciation. Both theoretical
developments and empirical studies over the past decades have increasingly
supported that sympatric speciation can occur in various cases
\citep{bolnick2007,bushsmith1998}. The process is often initiated by
ecological factors, such as differences in the use of resources or
habitats, which lead to divergent natural selection and the gradual
evolution of reproductive isolation between populations
\citep{schluter1998}. Cryptic sympatric speciation, where distinct lineages
are not morphologically distinguishable, may have also been overlooked in
the past, especially in the pre-genomic era \citep{jorde2018}. In the study
of diatoms, for instance, lineages of \textit{Pinnularia borealis} are
genetically distinct yet morphologically identical
\citep{pinseel2019}. Theoretical models have demonstrated that competition
for resources can naturally lead to new species in sympatry, as
specialization on different ecological niches reduces competition and
promotes ecological divergence \citep{dieckmann1999}. Recent genomic
approaches have revealed that the mechanisms of sympatric speciation are
more complex than previously assumed, involving interactions among
ecological, genetic, and behavioural factors \citep{foote2018}.

Mathematical theories can explain the possibility of sympatric and
parapatric speciation. We refer to the theoretical concept of symmetry
breaking, applied to speciation by \citet{golubitsky2003}. They
demonstrated that a system of coupled identical differential equations
describing some phenotype trait typically undergoes symmetry breaking into
differing phenotypes and applied it to Darwin's finches as an example. In
this study, we investigate the consequences of the Stewart-Golubitsky
approach for a three-clade speciation. Two subsequent speciation events in
the case of a slowly varying environment as well as a direct
differentiation into three clades in a fast changing environment are
discussed. Within the limits of the theory, we ask what kind of
differential equations would be needed to describe this situation, which
symmetry-reduced states are possible, and whether they are accessible from
a single species origin. Theoretical considerations are supported by
computer simulations on a specific model that can be generalized. Thus, our
extended model aims to provide a comprehensive framework that can be
applied to a range of organisms exhibiting sympatric or parapatric
speciation, thereby contributing to the broader understanding of this
evolutionary process and its modelling by the symmetry-breaking approach.

The paper is organised as follows.  The next section introduces the
structure of the considered models and some of their elementary properties.
In Section~\ref{sec:css}, we develop a numerical method for determining all
steady states of our models -- which is a non-trivial task, as the symmetry
leads to a very high number of steady states -- and analyse their
stability.  In Section~\ref{sec:numana}, we use this method for a closer
analysis of two specific model classes for a smaller number of populations
(up to ten).  We also perform a bifurcation analysis both with respect to
the environmental parameter and with respect to the coupling strength.  In
Section~\ref{sec:trajsim}, we analyse our models using numerical trajectory
simulations.  Main questions are the existence and accessibility of
three-clade states when the environmental parameter either jumps or is
ramped to a new value.  Finally, some conclusions are given.

\section{Three-clade model}

For the illustration of bifurcations into two species, \cite{gs:paper} used
the specific model
\begin{equation}
  \label{eq:Stewart model}
  \dot{x}_{k} = -x_{k}\Bigl(x_k^2-x_k-\mu\Bigr)-\sum_{j=1}^N x_j\,,
  \qquad k=1,\dots,N
\end{equation}
that we take in this slightly rewritten form as starting point for our considerations.
It is a special instance of the more general model class
\begin{equation}
  \label{eq:general}
  \dot{x}_{k} = f(x_k,\mu) +g(x_k, \mathbf{x}) 
\end{equation}
where we distinguish between a growth function $f$ describing the evolution
of the trait variable~$x_{k}$ of a population, following a general
nonlinear law, and a coupling term $g$ which connects the evolution of the
$N$ populations and in which $\mathbf{x}$ denotes the vector of
all~$x_{k}$.  The key requirement on $g$ is that the complete system
\eqref{eq:general} remains invariant under arbitrary permutations of the
populations, i.\,e.\ it possesses the full permutation group $S_{N}$ as
symmetry group.

\eqref{eq:general} is thus a system of $N$ identical differential equations
for the time derivative $\dot{x}_{k}$ of each single trait variable $x_{k}$
that characterizes a (sub)population $k$ (called by Stewart \textit{et al.}
"placeholder for organism development", POD). In this sense, population
stands here for a certain coarse-grained number of individuals or
strains. The trait variable could refer to a certain phenotype or
genotype. In the model \eqref{eq:Stewart model} of Stewart \textit{et al.},
$g(x_k, \mathbf{x})=-\sigma_{1}(\mathbf{x})=-(x_{1}+\cdots+x_{N})$
corresponds to the simplest, non-constant $S_{N}$-invariant coupling
(first-order coupling).  If for a certain value of the bifurcation
parameter $\mu$, representing a varying environmental parameter, a steady
state becomes unstable, the populations distribute over several states
generated by the bifurcation. Due to the coupling term, the specific values
$\mathbf{x}$ of a state depend on the $x$ values of all populations.
Symmetry breaking refers to the fact that solutions generally do not retain
the full permutation symmetry, thus subgroups (in mathematical and
biological sense) evolve that develop different traits, despite of
identical equations for each population of the species.  Stewart \textit{et
  al.} concentrated on the case that two different trait values emerge; we
will study mainly the case of three different traits.

At this point, some comments on the notation are appropriate. The vector
$\mathbf x$ characterizes the \textit{state of the system}. Its components
are the \textit{levels of the populations}. Dynamical equilibria with
$\dot{\mathbf{x}} = 0$ are called \textit{steady states}. In such a steady
state of the system, populations split into subsets each of which is
characterized by the same trait variable. We say that an \textit{$M$-fold
  split} or \textit{$M$-clade distribution} divides the total population
into $M$ non-empty subsets, each having a common level $x$.  Typically each
distribution has a different set of levels. We denote a distribution as an
ordered $M$-tuple of occupations $(N_{1},N_{2},\dots,N_{M})$ meaning that
$N_{1}$ populations occupy (exhibit) the lowest level, $N_{2}$~populations
the second lowest level and $N_{M}$ populations the highest level.  Here
$N_{1}+\cdots+N_{M}=N$, the total number of populations (and dimension of
the vector $\mathbf{x}$). If we know that levels are unoccupied, we write
\textit{extended distributions} by filling unoccupied levels with zeros.

We modify the original system in some respects: First, we want to allow the
simultaneous presence of three levels that we can identify with
clades. Considering only the uncoupled differential equations, choosing a
polynomial $f$ of degree three only leads to a maximum of two different
levels, basically because in a cubic equation three stationary solutions
are possible that alternate in stability (stability is obtained from the
sign of the first derivative at the roots of the polynomial, and this has
to change alternately). In the coupled system this is less obvious,
however, and we will have to determine the stability in a more complicated
procedure, as shown in a subsequent section.  But the obvious approach to
guarantee stable steady states with three different levels is the
transition to a polynomial $f$ of degree five.
 
We choose a \textit{specific} representation for this quintic polynomial,
namely by multiplying two quadratic polynomials with a linear function
(over the reals, any polynomial of degree five can be written in this
way). In each factor we include the environmental parameter $\mu$.  Thus,
we write
\begin{equation}
  \label{eq:fifth_order}
  f(x_k) = -(x_{k}-r\mu)(o_1 x_k^2+p_1 x_k+q_1 - \mu)(o_2 x_k^2+p_2 x_k+q_2 - \mu)\,.
\end{equation}
This representation of $f$ has the advantage that the steady states of the
uncoupled system can be easily calculated as roots by solving
$x_k-r \mu$=0, $o_1 x_k^2+p_1 x_k+q_1 - \mu$=0, and
$o_2 x_k^2+p_2 x_k+q_2 - \mu =0$. Below the bifurcation points, the roots
are imaginary and the number of real solutions drops. Biologically, we
reduce multiple decisions in which direction evolution can proceed, which
usually covers many traits, to a one-trait model, but couple these
decisions separately to the environment.

\begin{figure}[tb]
    \centering
    \includegraphics{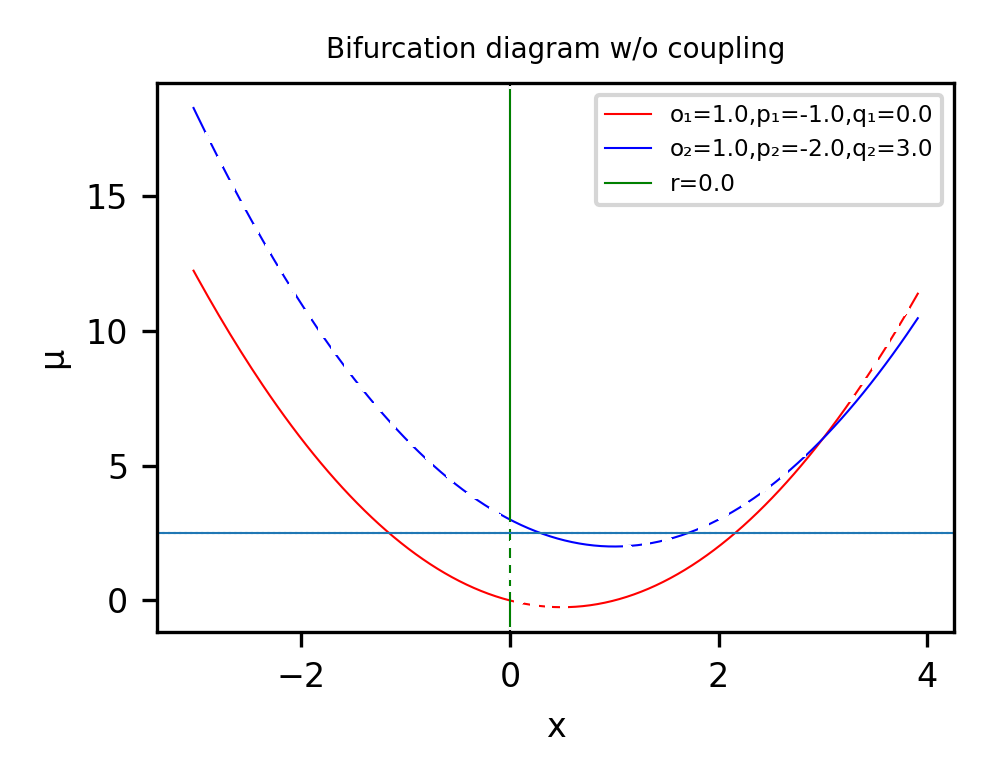}
    \caption{Bifurcation diagram without coupling. The steady levels are
      located on the green line and the red and blue parabolas (in colour
      image). Solid lines indicate stable branches, dashed lines unstable
      branches. The horizontal light blue line indicates a specific value
      of $\mu=2.5.$ for which five roots exist in the following order:
      stable (red), unstable (green), stable (blue), unstable (blue),
      stable (red).}
    \label{fig:BFD}
\end{figure}

In the bifurcation diagram (Fig.~\ref{fig:BFD}), in which these equations
are plotted, stable and unstable roots are indicated by green, red and blue
(or greyscale) lines for a specific system with the parameter values
$o_1=o_2=1, p_1=1, p_2=2, q_1=0, q_2=3, r=0$ that we will use for
demonstration throughout this paper. If $\mu$ is increased from negative to
positive values, first the same happens as in the original model of
eq.~\eqref{eq:Stewart model}, i.e.\ the starting steady state with all
$x_k=0$ becomes unstable at $(0,0)$ and a bifurcation to two states arises,
with a distribution that is determined by the initial stochastic
fluctuation. Later, at $\mu=2$, a third branch arises, and opens the
possibility of three-clade states. Stability of the branches is exchanged
whenever two curves are crossing (here at $(x,\mu)=(0,3)$ and $(3,6)$). In
this picture, the classic bifurcation types of one-dimensional dynamic
systems, \textit{saddle-node bifurcation}, \textit{transcritical
  bifurcation}, and \textit{pitchfork bifurcation} have some common origin,
namely the onset of a new bifurcation of saddle-node type at the vertex
points of the parabolas (here $(x_1,\mu_1)=(0.5,-0.25)$ and
$(x_2,\mu_2)=(1,2)$, respectively). A pitchfork bifurcation is constructed
as a special case by placing the vertex on some line of the other
solutions. We can identify the vertex points as a critical combination of
trait and environment, from which organism development can go into two
different directions. At intersections, leading to trans\-critical
behaviour with exchange of stability, the bifurcation occurs with a jump to
one branch apparently appearing ``out of the blue'' \citep{as:book}.

The possible bifurcations become more complicated, if the populations are
coupled. As coupling term in the general model \eqref{eq:general} we only
use the first elementary symmetric polynomial of the state vector
representing the sum influence of the coupled populations, but allow more
generally a polynomial in $x_k$ as prefactor
\begin{equation}
  \label{eq:genh}
  g(x_{k},\mathbf{x})=h(x_{k})\sigma_{1}(\mathbf{x})\,,
  \qquad k=1,\dots,N\,.
\end{equation}
Note that in this general form we impose no assumptions on the signs of the
intrinsic growth $f$ and the coupling $g$.

For numerical calculations and simulations, we consider only first and second order
coupling:
\begin{equation}
  \label{eq:coupling2nd}
  g(x_k, \mathbf{x}) = (\beta +\gamma x_{k}) \sigma_{1}\,.
\end{equation}
The latter seems appropriate for sexual reproduction mechanisms in which
different strains are crossed and the outcome is influenced by both
parental traits. $\beta$ and $\gamma$ correspond to the parameters $b_1$
and $c_2$, respectively, of Golubitsky and Stewart in their general
treatment of coupling up to third order \citep{golubitsky2003}. In their basic
model \eqref{eq:Stewart model}, $\beta = -1, \gamma=0$ is used. We will use
both parameters for tuning the coupling strength, but not simultaneously,
i.\,e.\ either $\beta$ or $\gamma$ is always set to zero.

A special characteristic of the general model inherited from the model of
Golubitsky and Stewart is the possibility of positive and negative values
for the levels. This has immediate consequences on the sign of coupling. If
only $\beta$ is taken into account, we find a kind of diffusive coupling,
expressed by the gradient of the level between different populations
$\beta (x_j-x_k)$ in which the $x_k$ term can be subsumed in the
function~$f$. Positive values for $\beta$ then mean a normal diffusion
process, aimed at balancing the values. The situation is different for the
second-order term $\gamma x_j x_k$. Depending on the specific signs of the
levels $x_j$ and $x_k$, and $\gamma$, the coupling can draw the levels in
positive or negative directions. Thus, the $\gamma$ term would lead to an
increase if both parental trait variables are equal in sign, and to a
decrease if they differ. This means that in a specific context, the model
has to be chosen carefully according to the direction a trait combination
is expected to occur.  For easy reference, the sign multiplication table is
given in Table~\ref{tab:sign}.

\begin{table}[tb]
    \centering
    \caption{Influence of sign of level and coupling parameters on time
      evolution. "$-$" means $g(x_k,\mathbf{x})<0$, i.e. level of state decreasing with time relative
      to uncoupled system), "$+$" $g(x_k,\mathbf{x})>0$,
      respectively.}
    \begin{tabular}{lcccc}
         \hline
         \textbf{coupling} & \textbf{($x_k<0, \sigma_1<0$)} & \textbf{($x_k>0, \sigma_1<0$)} &\textbf{($x_k<0, \sigma_1>0$)} &  \textbf{($x_k>0, \sigma_1>0$)}  \\
         \hline
         $\beta<0$ & $+$ & $+$ & $-$ & $-$\\
         $\beta>0$ & $-$ & $-$ & $+$ & $+$\\
         $\gamma<0$ & $-$ &  $+$ &  $+$ & $-$ \\
         $\gamma>0$ & $+$  & $-$  & $-$  & $+$ \\
          \hline
    \end{tabular}
    \label{tab:sign}
\end{table}

In our numerical standard model defined in the
bifurcation diagram in Fig.~\ref{fig:BFD}, positive and negative levels occur
for $\mu>0$, so all possible sign combinations are covered.  The stability
of a steady state in the coupled system is described by the Jacobian,
including the contributions of both $f(x_k)$ and $g(x_k,\mathbf{x})$, and
so the corresponding signs enter in the stability analysis as well.

Having introduced now the model that is suitable for the case of
three-clade speciation with a quintic polynomial and coupling terms
containing $\sigma_1$, we will address the following questions: In the
general or a specific model, how can we describe the steady states as
derived from the uncoupled system? How can we determine the stability of
specific distributions and see when a distribution loses its stability? Are
really all stable distributions accessible by either ramping the
bifurcation parameter or by jumping to a different value, describing a slow
or sudden change in environmental conditions, respectively?  We address
these questions algebraically whenever possible for the general model, and
numerically for the specific standard model. Specifically we are interested
in the stability and accessibility of three-clade states. The methods can
be easily adapted for other parameter values.

\section{Computing Steady States and Their Stability}\label{sec:css}

As mentioned above, we consider now the following class of models
\begin{equation}\label{eq:genmod}
  \dot{x}_{k} = f(x_{k})+h(x_{k})\sigma_{1}(\mathbf{x})\,,
  \qquad k=1,\dots,N\,,
\end{equation}
with $f$ and $h$ polynomials.  In this section, we will assume that the two
polynomials $f$ and $h$ do \emph{not} have a common, non trivial factor
$\psi$, i.\,e.\ there does not exist a polynomial $\psi$ of positive degree
such that $f(x)=\tilde{f}(x)\psi(x)$ and $h(x)=\tilde{h}(x)\psi(x)$.

Such a factor $\psi$ can be easily handled separately.  Its main effect is
the appearence of additional steady states.  Assume that a population
$x_{k}$ closely approaches a value $\bar{x}$ with $\psi(\bar{x})=0$.  Then
this population effectively ceases to evolve, as $\dot{x}_{k}$ will
approach zero and then remain zero (obviously, this is biologically not a
very meaningful behaviour).  Thus a search for steady states can be split
into two cases.  One first ignores the factor $\psi$ and computes all zeros
of the system $\tilde{f}(x_{k})+\tilde{h}(x_{k})\sigma_{1}(\mathbf{x})=0$
with $k=1,\dots,N$.  Then one considers subsequently all subsets
$L\subseteq\{1,\dots,N\}$ and sets for each $\ell\in L$ the population
$x_{\ell}$ to a value $\bar{x}_{\ell}$ with $\psi(\bar{x}_{\ell})=0$ and
determines the solutions of the system
$$\tilde{f}(x_{k})+\tilde{h}(x_{k})\Bigl(\sum_{i\notin
  L}x_{i}+\sum_{\ell\in L}\bar{x}_{\ell}\Bigr)=0\,,\qquad k\notin L\,.$$
Obviously, for larger $N$ and factors $\psi$ with a larger number of zeros,
this leads to a combinatorial explosion of the number of computations to be
done. 

\subsection{Finding Steady States}

In a similar manner as just discussed for zeros of a common factor $\psi$,
it suffices for finding all steady states to restrict our attention to
steady states $\bar{\mathbf{x}}$ where all components $\bar{x}_{k}$ are
non-zero.  Indeed, assume that exactly $n<N$ components of
$\bar{\mathbf{x}}$ were zero; then the remaining components define a steady
state without zero components of a system of the form \eqref{eq:genmod},
but with size $\bar{N}=N-n$.  Hence, the steady states with zero components
can be found by iteratively studying the cases $\bar{N}=1,2,3,\dots,N-1$.

For finding all steady states of \eqref{eq:genmod} (independent of their
stability properties), we need all \emph{real} solutions of a system of $N$
polynomial equations in $N$ variables $x_{k}$.  Already for moderate values
of $N$ and rather low degrees of the polynomials~$f$ and~$g$, this task
requires very substantial computations which cannot be performed
symbolically.  Because of the $S_{N}$-symmetry of the system, we must also
expect a very large number of steady states entailing that even a purely
numerical computation is non-trivial (in particular, if one wants to ensure
that no steady state is overlooked).  We now describe an alternative
approach leading to univariate computations that are much simpler and can
be easily performed numerically.

We are interested in \emph{symmetry breaking} steady states.  This means
that we consider a (fixed) $M$-fold distribution
$\mathcal{N}=(N_{1},\dots,N_{M})$.  With each such distribution
$\mathcal{N}$, we associate a \emph{reduced} system of dimension~$M$
\begin{equation}
  \label{eq:redgensys}
    \dot{y}_{i} = f(y_{i})+h(y_{i})\sum_{j=1}^{M}N_{j}y_{j}\,,
  \qquad i=1,\dots,M\,,
\end{equation}
describing the evolution of the levels~$y_{i}$.  This system arises when we
assume that in \eqref{eq:genmod} the components $x_{k}$ are partitioned
into $M$ subsets of sizes $N_{1},\dots,N_{M}$ and that all components in
the $i$th subset take the same value $y_{i}$ (because of the $S_{N}$
invariance, it does not matter which components are in which partition).
The original $S_{N}$-symmetry of \eqref{eq:genmod} is thus broken into an
$S_{N_{1}}\times\cdots\times S_{N_{M}}$-symmetry.

All steady states belonging to a fixed distribution
$\mathcal{N}=(N_{1},\dots,N_{M})$ can be determined in the following
manner.  Each such steady state has $N_{i}$~components equal to some value
$y_{i}$ where the $y_{i}$ are pairwise distinct.  Since we assume that the
steady state $\bar{\mathbf{x}}$ has no non-zero components, all the $y_{i}$
are also non-zero and the conditions for a steady state can be written as
\begin{equation}\label{eq:geneqcond}
  \frac{f(y_{1})}{h(y_{1})}=\cdots=\frac{f(y_{M})}{h(y_{M})}=
  -\bigl(N_{1}y_{1}+\cdots+N_{M}y_{M}\bigr)\,.
\end{equation}
Note that it cannot happen that $h(y_{j})=0$ for some~$j$.  For a steady
state, we must then also have $f(y_{j})=0$ and thus $y_{j}$ would be a
common zero of $f$ and $h$.  But this cannot happen, as we assumed that $f$
and $h$ have no non trivial common factor.

Consider the graph of the rational function $\rho=f/h$.  By determining the
poles and the local minima and maxima of $\rho$, we can identify sets
$\mathcal{I}_{M'}\subset\RR$ (which are either empty or finite unions of
open intervals and isolated points) such that for each
$c\in \mathcal{I}_{M'}$ the equation $\rho(x)=c$ has exactly $M'\geq M$
distinct solutions $\bar{y}_{1}<\cdots<\bar{y}_{M'}$ (counted
\emph{without} multiplicity) which represent the possible levels.  We
associate with such a solution set and the given distribution
$\mathcal{N}$ an \emph{extended distribution}
$\mathcal{N}'=(N'_{1},\dots,N'_{M'})$ of length $M'$ such that $M'-M$
entries vanish and each of the remaining $M$ entries takes a different
value $N_{i}$.  Then we define an auxiliary function
$\varphi_{\mathcal{N}}\colon\mathcal{I}_{M'}\to\RR$ by setting
$\varphi_{\mathcal{N}}(c)=c+\sum_{j=1}^{M'}N'_{j}\bar{y}_{j}$ (recall that
the levels $\bar{y}_{j}$ also depend on $c$).

Since the levels $\bar{y}_{j}$ are roots of a rational equation,
$\varphi_{\mathcal{N}}$ is an algebraic and thus in particular a continuous
and almost everywhere differentiable function.  We must determine its zeros
in $\mathcal{I}_{M'}$.  Because of the residual
$S_{N_{1}}\times\cdots\times S_{N_{M}}$-symmetry, each zero $\bar{c}$
defines a whole family of steady states of \eqref{eq:genmod} containing
$\binom{N}{N_{1}\ N_{2}\ \cdots\ N_{M}}$\footnote{The multinomial
  coefficient $\binom{N}{N_{1}\ N_{2}\ \cdots\ N_{M}}$ is defined as
  $N!/\prod_{i=1}^{M}(N_{i}!)$.} different steady states obtained by
permuting its coordinates.  Indeed, take those levels $\bar{y}_{j}$ with
$\rho(\bar{y}_{j})=\bar{c}$ such that $N'_{j}>0$ and assign it to $N'_{j}$
coordinates of a point $\bar{\mathbf{x}}\in\RR^{N}$.  Then
$\bar{\mathbf{x}}$ is a steady state of the original system
\eqref{eq:genmod} and properly ordered these $M$ levels $\bar{y}_{j}$
define a steady state of the reduced system \eqref{eq:redgensys}.

Our approach requires only solving univariate equations for which many
numerical methods are available.  Computing the levels for a given
value~$c$ is a purely polynomial computation.  Determining the zeros of
$\varphi_{\mathcal{N}}$ is preferably done with a derivative-free numerical
method, since we do not have a closed-form representation of
$\varphi_{\mathcal{N}}$.

Generically, $\varphi_{\mathcal{N}}$ can have only finitely many, isolated
zeros in a set $\mathcal{I}_{M'}$.  Our method is equivalent to solving the
$(M+1)$-dimensional nonlinear system
\begin{equation}
  \begin{gathered}
    \rho(y_{i})+c=0\,,\qquad i=1,\dots,M\,,\\
    \bigl(N_{1}y_{1}+\cdots+N_{M}y_{M}\bigr)-c=0\,,
  \end{gathered}
\end{equation}
but exploits that the first $M$ equations are univariate and only in the
last equation a (linear) coupling of the solutions happens.  We are only
interested in solutions where the $y_{i}$ are pairwise distinct.  If we
furthermore assume that no $y_{i}$ belongs to a critical point of $\rho$
(which generically is the case), then the Jacobian of this system is
non-singular.  Hence, by the Inverse Function Theorem, the left hand side
of the system defines generically a locally invertible function and if a
solution exists, then it is locally unique, i.\,e.\ isolated.  Since we are
dealing with polynomials, only finitely many solutions are possible.  In
our experiments, we observed that if solutions exist, then they are always
unique.  But a rigorous proof of this fact seems to be hard.

Finding \emph{all} steady states of \eqref{eq:genmod} has now been divided
into a large number of small univariate problems.  However, a combinatorial
explosion in the number of these small problems occurs for larger values
of~$N$.  By the ``stars and bars'' formula, there are
$C_{N,M'}=\binom{N+M'-1}{N}$ possibilities to distribute $N$ identical
balls into $M'$ distinct boxes.  Hence, $C_{N,M'}$ describes the number of
extended distributions that must be considered when searching for all
steady states corresponding to a split into up to $M'$ distinct levels
(provided $\mathcal{I}_{M'}$ is non-empty).  Considered as a function of
$N$, the binomial coefficient $C_{N,M'}$ is a polynomial of degree $M'-1$
and we obtain a polynomial complexity of the problem.  We will be mainly
concerned with the cases $M'=2$ and $M'=4$ because of the specific class of
models which we will study. In Section~\ref{sec:ststN10}, it will be shown
for a specific model that the number of steady states typically grows
exponentially so that our approach very quickly becomes more efficient than
any attempt to determine the steady states directly.

\subsection{Stability Analysis}
\label{sec:ssss}

For our purposes, only asymptotically stable steady states are of interest.
Hence, as next step, we must analyse the stability of the found steady
states.  Let $\bar{\mathbf{x}}$ be an arbitrary steady state of
\eqref{eq:genmod} (possibly containing zero components).  We want to
determine its stability using the Jacobian of \eqref{eq:genmod}.  We assume
that $\bar{\mathbf{x}}$ belongs to an $M$-fold split $(N_{1},\dots,N_{M})$
with corresponding values $\bar{y}_{i}$ for $i=1,\dots,M$.  We introduce
for each index~$i$ the following two quantities
\begin{equation}\label{eq:ab}
  a_{i}=h(\bar{y}_{i})\,,\qquad b_{i}=f'(\bar{y}_{i})+h'(\bar{y}_{i})c\;,
\end{equation}
where as before  $c=-\sum_{j=1}^{M}N_{j}\bar{y}_{j}$.

In a straightforward calculation, one finds that the Jacobian of
\eqref{eq:genmod} at the equilibrium $\bar{\mathbf{x}}$ has the following
block form\footnote{Permutations of the coordinates of $\bar{\mathbf{x}}$
  lead only to simultaneous permutations of rows and columns of $J$ which
  do not affect the determinant.  Thus the following analysis is valid for
  the whole class of steady states belonging to the distribution
  $(N_{1},\dots,N_{M})$, as the chosen extended distribution
  $(N'_{1},\dots,N'_{M'})$ only affects which values $\bar{y}_{i}$ appear.}
\begin{equation}\label{eq:Jblock}
  J=
  \begin{pmatrix}
    J_{1} & A_{12} & \cdots & \cdots & A_{1M} \\
    A_{21} & J_{2} & A_{23} & \cdots & A_{2M} \\
    & & \ddots & & \\
    & & & \ddots & \\
    A_{M1} & A_{M2} & \cdots & A_{M,M-1} & J_{M}
  \end{pmatrix}
\end{equation}
with submatrices $J_{k}\in\RR^{N_{k}\times N_{k}}$ and
$A_{k\ell}\in\RR^{N_{k}\times N_{\ell}}$ given by
\begin{equation}\label{eq:JA}
  J_{k}=
  \begin{pmatrix}
    b_{k}+a_{k} & & a_{k} \\
    & \ddots & \\
    a_{k} & & b_{k}+a_{k}
  \end{pmatrix}\,,\qquad
  A_{k\ell}=
  \begin{pmatrix}
    a_{k} & \cdots & a_{k} \\
    \vdots & & \vdots \\
    a_{k} & \cdots & a_{k}
  \end{pmatrix}\,.
\end{equation}

Exploiting this structure, we can compute in a few steps detailed in
Appendix~\ref{sec:lac} the characteristic polynomial $\chi_{J}(\lambda)$ of
the Jacobian~$J$ obtaining it in the following factored form
\begin{equation}\label{eq:charJ}
  \chi_{J}(\lambda)=
  \prod_{j=1}^{M}(b_{j}-\lambda)^{N_{j}-1}
  \chi_{J_{\mathrm{red}}}^{(M)}(\lambda)
\end{equation}
where $J_{\mathrm{red}}^{(M)}\in\RR^{M\times M}$ is the Jacobian of the
reduced system \eqref{eq:redgensys} evaluated at the corresponding
equilibrium $\bar{\mathbf{y}}$ (see \eqref{eq:Jred} for an explicit
expression). 

This factorisation of the characteristic polynomial $\chi_{J}$ possesses a
simple interpretation.  We are looking at steady states with a residual
$S_{N_{1}}\times S_{N_{2}}\times\cdots\times S_{N_{M}}$-symmetry.  For each
natural action $\alpha$ of this symmetry group, there exists an invariant
subspace $V_{\alpha}\subset\RR^{N}$ such that all trajectories defined by
an initial point on $V_{\alpha}$ remain on $V_{\alpha}$ and thus also
possess an
$S_{N_{1}}\times S_{N_{2}}\times\cdots\times S_{N_{M}}$-symmetry.  The
reduced system \eqref{eq:redgensys} describes the dynamics on $V_{\alpha}$
and hence the eigenvalues of $J_{\mathrm{red}}^{(M)}$ decide the stability
of the steady states for the reduced dynamics on $V_{\alpha}$.  The
remaining eigenvalues $b_{1},\dots,b_{M}$ of the full Jacobian~$J$ describe
in a neighbourhood of the steady state the dynamics transversal to the
subspace $V_{\alpha}$.

Since for each index $j$ with $N_{j}>0$ we find $b_{j}$ as a real
eigenvalue (of algebraic multiplicity $N_{j}-1$), a \emph{necessary}
condition for asymptotic stability is $b_{j}<0$ for all $j$ with $N_{j}>0$.
To obtain also a \emph{sufficient} condition, we must in addition require
that all eigenvalues of $J_{\mathrm{red}}^{(M)}$ have a negative real part,
i.\,e.\ that $\bar{\mathbf{y}}$ is an asymptotically stable steady state of
the reduced system~\eqref{eq:redgensys}.  In this statement, we have
implicitly assumed that $\bar{\mathbf{x}}$ is a hyperbolic steady state,
i.\,e.\ that no eigenvalue has a vanishing real part.  The analysis of
non-hyperbolic steady states is much harder and cannot be performed at this
level of generality (in our numerical computations we never encountered
such a situation, but we cannot exclude that they exist).

$\chi_{J_{\mathrm{red}}}^{(M)}(\lambda)$ is a polynomial of degree~$M$ that
is difficult to analyse for larger values of~$M$.  We are particularly
interested in the cases $M=2$ and $M=3$ where we assume that all $N_{j}>0$.
For $M=2$, it is straightforward to find
\begin{equation}\label{eq:charJred2}
  \chi_{J_{\mathrm{red}}}^{(2)}(\lambda)=
  \lambda^{2} -\bigl(N_{1}a_{1}+N_{2}a_{2}+b_{1}+b_{2}\bigr)\lambda +
  \bigl(b_{1}b_{2}+N_{1}a_{1}b_{2}+N_{2}a_{2}b_{1}\bigr)\,.
\end{equation}
According to the \emph{Routh--Hurwitz criterion}, all zeros of a real
quadratic polynomial $P(\lambda)=\lambda^{2}+c_{1}\lambda+c_{0}$ have a
negative real part, if and only if both $c_{1}$ and $c_{0}$ are positive.
Hence, we obtain as a necessary and sufficient condition for
\emph{asymptotic} stability in the case $M=2$:
\begin{equation}\label{eq:stab2}
  \begin{aligned}
    b_{1} &< 0\,,\\
    b_{2} &< 0\,,\\
    N_{1}a_{1}+N_{2}a_{2}+b_{1}+b_{2} &< 0\,,\\
    -\bigl(b_{1}b_{2}+N_{1}a_{1}b_{2}+N_{2}a_{2}b_{1}\bigr) &< 0\,.
  \end{aligned}
\end{equation}

For $M=3$, a tedious but still straightforward calculation yields
\begin{equation}\label{eq:charJred3}
  \begin{aligned}
    -\chi_{J_{\mathrm{red}}}^{(3)}(\lambda) & = 
    \lambda^{3} -
    \bigl(N_{1}a_{1}+N_{2}a_{2}+N_{3}a_{3}+b_{1}+b_{2}+b_{3}\bigr)\lambda^{2} \\
    &  {}+\Bigl(b_{1}b_{2}+b_{1}b_{3}+b_{2}b_{3}+{} \\
    & \phantom{{}-{}\bigl(} {} + (b_{2}+b_{3})N_{1}a_{1}+(b_{1}+b_{3})N_{2}a_{2}+
    (b_{1}+b_{2})N_{3}a_{3}\bigr)\lambda \\
    &  {}- \bigl(b_{1}b_{2}b_{3} + b_{2}b_{3}N_{1}a_{1} +
    b_{1}b_{3}N_{2}a_{2} + b_{1}b_{2}N_{3}a_{3}\bigr)\,.
  \end{aligned}
\end{equation}
According to the Routh--Hurwitz criterion, all zeros of a real cubic
polynomial $P(\lambda)=\lambda^{3}+c_{2}\lambda^{2}+c_{1}\lambda+c_{0}$
have a negative real part, if and only if all three coefficients $c_{i}$
are positive and in addition $c_{1}c_{2}>c_{0}$.  Thus, a neccessary and
sufficient condition for asymptotic stability in the case $M=3$
is:
\begin{equation}\label{eq:stab3}
  \begin{aligned}
    b_{1} &< 0\,,\\
    b_{2} &< 0\,,\\
    b_{3} &< 0\,,\\
    -c_{2}=N_{1}a_{1}+N_{2}a_{2}+N_{3}a_{3}+b_{1}+b_{2}+b_{3} &< 0\,,\\
    -c_{1}=-\bigl(b_{1}b_{2}+b_{1}b_{3}+b_{2}b_{3}+{} \phantom{(b_{1}+b_{3})N_{2}a_{2}+
      (b_{1}+b_{2})N_{3}a_{3}} \\
    {} + (b_{2}+b_{3})N_{1}a_{1}+(b_{1}+b_{3})N_{2}a_{2}+
     (b_{1}+b_{2})N_{3}a_{3}\bigr) &< 0\,,\\
     -c_{0}=b_{1}b_{2}b_{3} + b_{2}b_{3}N_{1}a_{1} +
     b_{1}b_{3}N_{2}a_{2} + b_{1}b_{2}N_{3}a_{3} &< 0\,,\\
     c_{0}-c_{1}c_{2} &<0\,.
  \end{aligned}
\end{equation}

\subsection{Two Special Classes of Models}\label{sec:scm}

For further analysis, we restrict to two special classes of models.  In
both classes, $f$ is assumed to be of the form
\begin{equation}
  f(x)=-xs(x)
\end{equation}
where $s(x)$ is a polynomial with a positive leading coefficient.  In the
notation of \eqref{eq:fifth_order}, this means that we choose $r=0$.  For
the coupling term, we consider the two simplest possibilities:
\begin{equation}
  h_{1}(x)=\beta\,,\qquad h_{2}(x)=\gamma x\,.
\end{equation}
The first variant corresponds to a first-order coupling.  In the second
variant, we have a pure second-order coupling and a common factor
$\psi(x)=x$ of the polynomials $f$ and $h_{2}$ arises.  As in this special
case, $0$ is the only zero of the common factor, the handling of the common
factor is equivalent to the handling of steady states with zero components
and can be easily performed by analysing also lower dimensional versions of
our model. 

These two variants are not only natural to consider from an application
point of view.  In both cases the determination of the steady states
considerably simplifies, as no rational functions have to be considered: in
the first case, the denominator becomes a constant; in the second case, we
can simply ignore the common factor~$x$.  This is special for the case
$r=0$ which has the feature that one root of $f$ remains fixed at
zero. Therefore, a population which attained this value cannot evolve
anymore.  Depending on the situations to which the model is applied, this
might be useful, or it can be avoided by assigning a non-zero value to~$r$.
Such a non-zero value would also prevent a stable state for $\mu=0$
becoming stable again at higher $mu$ values according to
Fig.~\ref{fig:BFD}.

\subsubsection{First-Order Coupling}

The conditions (\ref{eq:geneqcond}) for a steady state take for a
first-order coupling the simpler, purely polynomial form
\begin{equation}\label{eq:eqcond1}
  f(y_{1})=\cdots=f(y_{M})=
  -\beta\bigl(N_{1}y_{1}+\cdots+N_{M}y_{M}\bigr)\,.
\end{equation}
The quantities $a_{i}$ and $b_{i}$ deciding about the stability of the
steady states are easily determined as
\begin{equation}\label{eq:abfoc}
  a_{i}=\beta\,,\qquad b_{i}=f'(\bar{y}_{i})\;.
\end{equation}
Hence all the $a_{i}$ are simply given by the coupling constant~$\beta$ and
independent of the considered level.  The $b_{i}$ are the slopes of the
function $f$ at the levels~$\bar{y}_{i}$.

Recall that a necessary condition for asymptotic stability is that all the
$b_{i}$ are negative.  Since $f$ is assumed to be a polynomial, the signs
of the slopes alternate between neighbouring possible levels $y_{j}$.  This
trivial observation implies that only every second $y_{j}$ can be used as a
level $\bar{y}_{i}$ of a stable steady state.  If we assume that $s$ is a
polynomial of even degree with a positive leading coefficient, then
$f(x)=-xs(x)$ is a polynomial of odd degree with a negative leading
coefficient.  If we choose $c$ such that $f^{-1}(c)$ contains $\deg{f}$
elements $y_{1}<y_{2}<\cdots<y_{\deg{f}}$, then $f'(y_{1})<0$ and hence the
same is true for all $y_{i}$ with an odd index.  For the existence of an
extended $M$-fold distribution leading to an asymptotically stable steady
state we thus need that $\deg{f}\geq 2M-1$.  For the case $M=3$
mainly studied by us, we therefore need $\deg{f}\geq5$.

If $\beta<0$, then it is easy to see that the neccessary conditions
$b_{i}<0$ automatically imply the remaining conditions in \eqref{eq:stab2}
and \eqref{eq:stab3}, as these are then sums of only negative summands.  If
the coupling constant~$\beta$ is positive, then it must be relatively small
compared to the absolute values of the slopes~$b_{i}$ for a stable steady
state.  For example, for $M=2$ we obtain the two conditions
\begin{equation}
  -(b_{1}+b_{2})>N\beta\,,\qquad b_{1}b_{2}>-(N_{1}b_{2}+N_{2}b_{1})\beta\,.
\end{equation}
They can always be satisfied by rescaling $f$ with a sufficiently large
positive factor~$q$, i.\,e.\  by replacing $f(x)$ with $qf(x)$.  Indeed,
then all $b_{i}$ also get multiplied by $q$ and in the above inequalities
we always have on the left hand side a higher degree in the $b_{i}$ than on
the right hand side.  This also remains true for $M=3$, as one can easily
see from \eqref{eq:stab3}.

\subsubsection{Second-Order Coupling}

The conditions (\ref{eq:geneqcond}) for a steady state take for a
second-order coupling the simpler, purely polynomial form
\begin{equation}\label{eq:eqcond2}
  s(y_{1})=\cdots=s(y_{M})=\gamma\bigl(N_{1}y_{1}+\cdots+N_{M}y_{M}\bigr)\,.
\end{equation}
Thus compared to the first-order case, we can work with the polynomial~$s$
instead of~$f$, i.\,e.\ at a lower degree.  The quantities $a_{i}$ and
$b_{i}$ deciding about the stability of the steady states can again be
easily interpreted.  We find
\begin{equation}\label{eq:absoc}
  a_{i}=\gamma \bar{y}_{i}\,,\qquad b_{i}=-\bar{y}_{i}s'(\bar{y}_{i})\;.
\end{equation}
Thus the $a_{i}$ are determined by the coupling constant~$\gamma$
multiplied with the corresponding levels $\bar{y}_{i}$ and the $b_{i}$ are
the negative product of the levels $\bar{y}_{i}$ with the slopes
$s'(\bar{y}_{i})$ of the function~$s$ at the levels.  A necessary
condition for stability is hence that the occupied levels and the slopes
there have always the same sign.

This observation allows already some conclusions.  For $M=2$, we may
consider models where $s$ is a polynomial of degree~$2$ (like in the
original model of Golubitsky and Stewart) and thus only two levels are
possible.  If the two levels have opposite signs (which will always be the
case, if the minimum of $s$ is sufficiently close to $x=0$), then $b_{1}$
and $b_{2}$ will both be negative.  Indeed, the slope at the negative level
will be negative and at the positive level it will be positive.

For $M=3$, one may hope that it sufficed if $s$ was a cubic polynomial, but
it is easy to see that in this case no stable three-clade state can exist.
If all three levels have the same sign, then at at least one of them the
slope must be of opposite sign.  If one level has a different sign than the
other two, then at the two levels with the same sign the slopes will be of
different signs implying that at least one $b_{i}$ is positive.

Thus for $M=3$, we need at least a quartic polynomial as~$s$ for which the
region $\mathcal{I}_{4}$ is non-empty.  For $\deg{s}=4$, let
$y_{1}<y_{2}<y_{3}<y_{4}$ be the four possible levels.  We have to choose
three of them as levels $\bar{y}_{1}$, $\bar{y}_{2}$, $\bar{y}_{3}$.  It is
straightforward to analyse the different possibilities and to discover that
an asymptotically stable steady state can arise in only two ways.  The
first possibility is $y_{1}<0<y_{2}<y_{3}<y_{4}$ and the choice
$\bar{y}_{1}=y_{1}$, $\bar{y}_{2}=y_{2}$, $\bar{y}_{3}=y_{4}$.  The second
possibility is $y_{1}<y_{2}<y_{3}<0<y_{4}$ and the choice
$\bar{y}_{1}=y_{1}$, $\bar{y}_{2}=y_{3}$, $\bar{y}_{3}=y_{4}$.  Thus one
level must have a different sign than the other three and one of the inner
possible levels must be left out.  Extending these considerations, it is
easy to see that for any $M>2$ the polynomial $s$ must have at least degree
$M+1$.

\section{Numerical Analysis for Small~$N$}
\label{sec:numana}

\begin{figure}[tb]
  \centering
  \begin{subfigure}[t]{0.45\textwidth}
    \includegraphics[width=\textwidth]{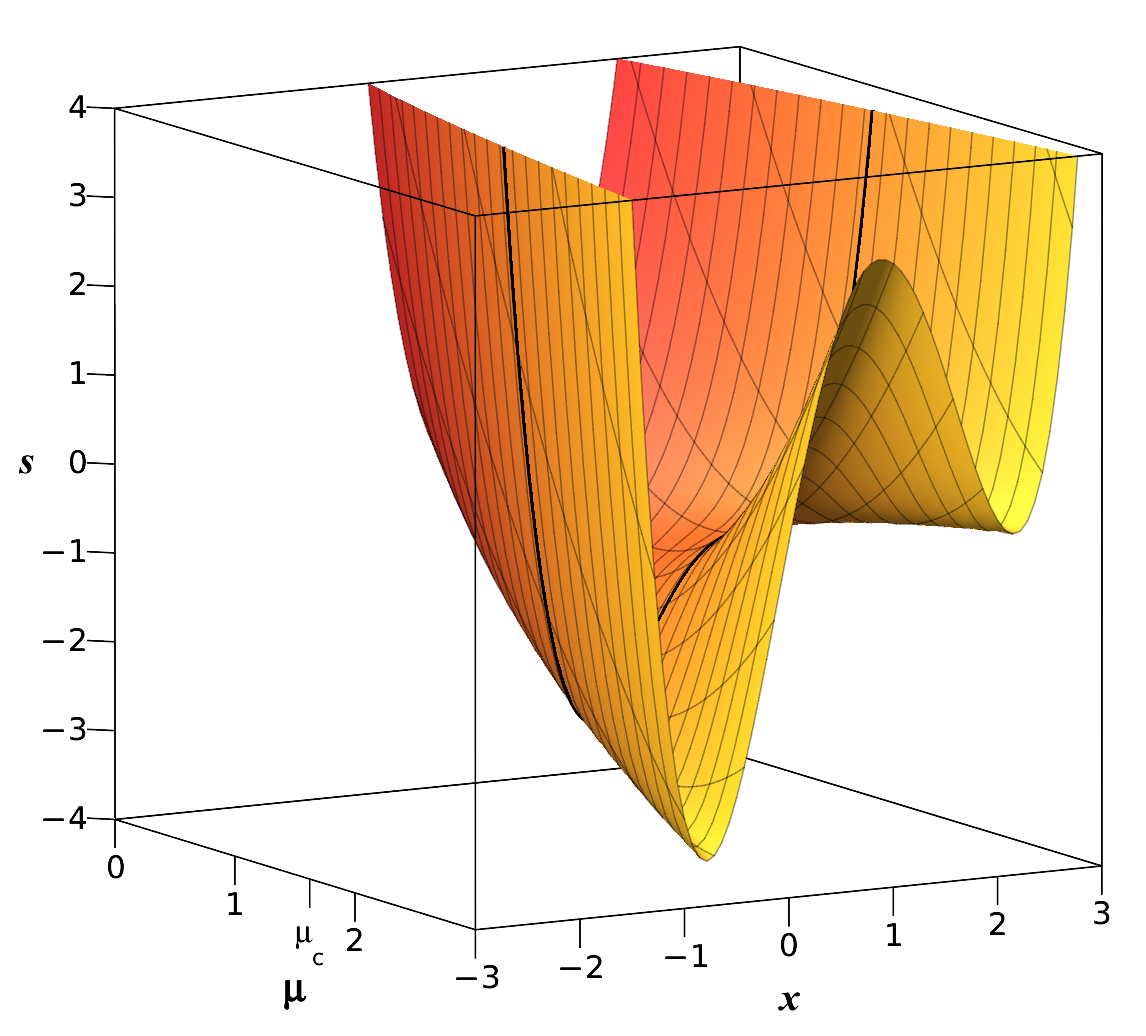}
    \caption{Dependency of the shape of $s$ on the environmental parameter $\mu$.}
    \label{fig:shapes}
  \end{subfigure}%
  \hfill
  \begin{subfigure}[t]{0.45\textwidth}
    \includegraphics[width=0.9\textwidth]{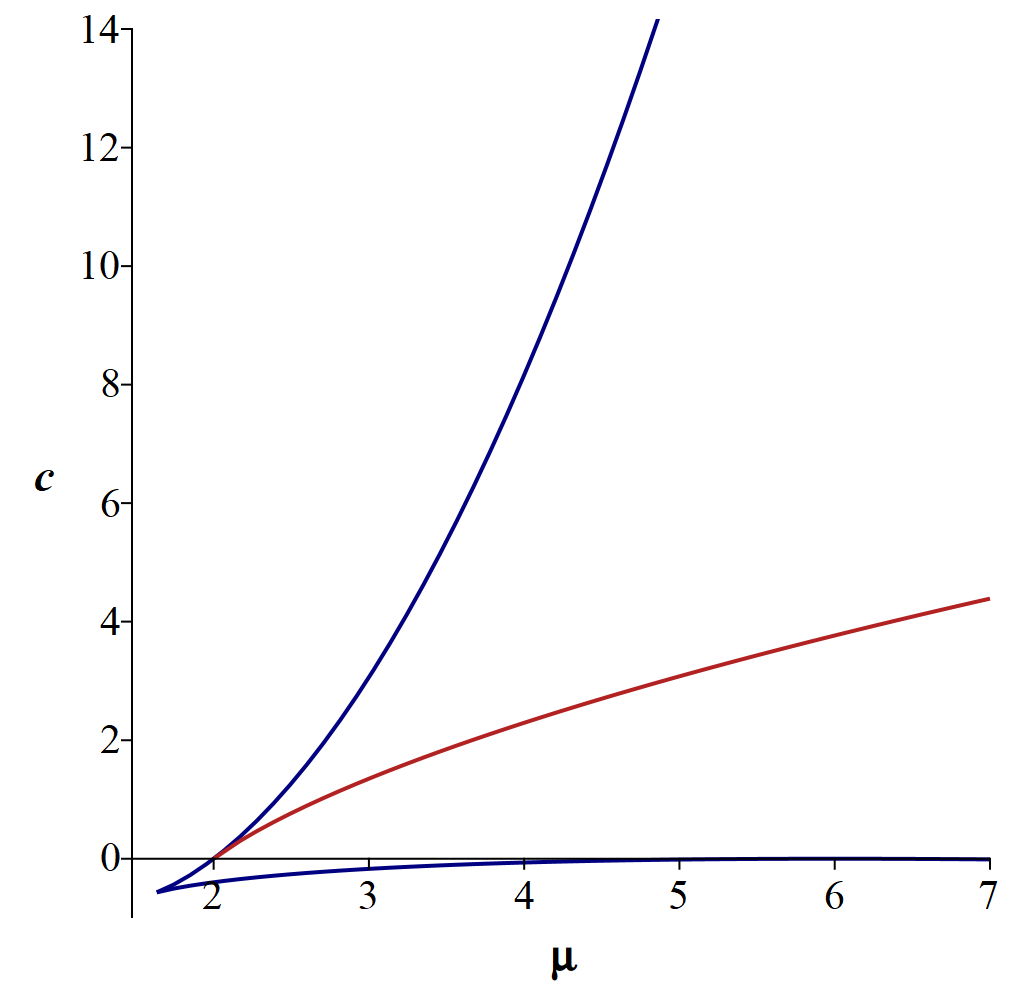}
    \caption{Range of possible $c$-values and the critical $c$-value for
      the extended three-clade distribution $(3,1,0,1)$.}
    \label{fig:ccrit}
  \end{subfigure}
  \caption{Solvability of \eqref{eq:eqcond2}.}
  \label{fig:solvcond}
\end{figure}

For the remainder of this paper, we will work throughout with the specific
polynomial
\begin{equation}\label{eq:fs}
  f(x)=-xs(x)\,,\qquad s(x)=(x^{2}-x-\mu)(x^{2}-2x+3-\mu)
\end{equation}
which was already used for the bifurcation diagram in Fig.~\ref{fig:BFD}.
Fig.~\ref{fig:shapes} shows the graph of $s(x)$ for values of the parameter
$\mu$ between $0$ and $3$.

One sees that only from a critical value $\mu_{c}$ on a non-empty region
$\mathcal{I}_{4}$ with four preimages exists; for lower values only two
preimages can be found.  It is straightforward to determine this critical
value via elementary calculus: we consider $s$ as a function of $x$ and
$\mu$ and then solve the system
\begin{equation}
  \label{eq:muc}
  \frac{\partial}{\partial x}s(x,\mu)=0\,,\qquad
  \frac{\partial^{2}}{\partial x^{2}}s(x,\mu)=0
\end{equation}
(describing a critical point which generically is not an extremum) for $x$
and~$\mu$.  As \eqref{eq:muc} is cubic in $x$, we solved it numerically and
obtained $\mu_{c}\approx1.62377$ (and $x\approx1.27002$).  In
Fig.~\ref{fig:shapes}, the graph of $s$ for the specific value
$\mu=\mu_{c}$ is marked as a thick black line.

For the steady states of the uncoupled system (i.\,e.\  for
$\beta=\gamma=0$), it is decisive how many zeros the function $f$
possesses.  Obviously, $x=0$ is always a zero.  Further zeros are provided
by $s$.  The number of zeros of $s$ changes, when an extremum crosses the
$x$-axis.  Hence the critical values of $\mu$ can be determined by solving
the system
\begin{displaymath}
  \frac{\partial}{\partial x}s(x,\mu)=0\,,\qquad s(x,\mu)=0\,.
\end{displaymath}
In this case, the three solutions can be computed exactly (and in fact noted
in Fig.~\ref{fig:BFD}).  Again the $x$-values are of no interest and
changes happens at $\mu_{1}=-1/4$, $\mu_{2}=2$ and $\mu_{3}=6$ (at these
values $s$ has one double zero).  For $\mu<\mu_{1}$, $s$ has no zeros, for
$\mu=\mu_{1}$ one zero, for $\mu_{1}<\mu<\mu_{2}$ two zeros, for
$\mu=\mu_{2}$ three zeros, for $\mu_{2}<\mu<\mu_{3}$ four zeros, for
$\mu=\mu_{3}$ three zeros and for $\mu>\mu_{3}$ again four zeros.  For
counting the zeros of $f$, one must also take into account the two special
cases $\mu=0$ and $\mu=3$ for which $x=0$ is a zero of $s$ and hence a
double zero of $f$.  In Fig.~\ref{fig:BFD}, $\mu_{1}$ defines the minimum
of the red parabola and $\mu_{2}$ the one of the blue parabola.  At
$\mu_{3}$, the red and the blue parabola intersect.  The two special cases
$0$ and $3$ mark where one of the parabolas intersects the green line.

As we are mainly interested in three-clade states, we now indicate how the
existence of such states for the above $f$ can be studied using the method
developed in Sect.~\ref{sec:css}.  As already discussed in
Sect.~\ref{sec:scm}, for obtaining a split into three levels, the
polynomial~$s$ must have at least degree~$4$ -- which is the case for our
choice -- and we need a region $\mathcal{I}_{4}$ of $c$-values where there
are exactly four preimages in the set $s^{-1}(c)$.  But even then the
existence of a solution of the system \eqref{eq:eqcond2} is not guaranteed.

Given a value $\mu\geq\mu_{c}$, one must study the solvability of
\eqref{eq:eqcond2} separately for each extended distribution
$(N_{1}',\dots,N_{4}')$ where exactly one $N_{i}'$ is zero, i.\,e.\ we must
pick three of the four preimages and choose how often each of them should
appear as a component of the desired steady state.  Fig.~\ref{fig:ccrit}
depicts the solvability of \eqref{eq:eqcond2} for $N=5$ and the extended
distribution $(3,1,0,1)$.  The blue lines bound the range of $c$-values for
which four preimages exists, i.\,e.\ the region $\mathcal{I}_{4}$ for each
value of $\mu$; the red line gives the critical $c$-value corresponding to
a steady state.  As one can clearly see, the red line intersects the upper
boundary meaning that only from a $\mu$-value above $2$ indeed a solution
exists for this extended distribution.  The plot also indicates that
probably for all $\mu>2$ a solution exists.

While the blue lines are always the same for any distribution, as they are
determined by the growth function~$s$ and its dependency on the
environmental parameter~$\mu$, the red line will be different for each
case.  For some extended distributions, no solutions may exist at all or
solutions may exist only in a finite $\mu$-range.  In Sect.~\ref{sec:N5}
below, we will study for $N=5$ the different possibilities in detail taking
also the stability of the steady states into account.

\subsection{Finding All Steady States for $N\leq10$ and Second-Order
  Coupling}\label{sec:ststN10} 

We demonstrate now the efficiency of our approach to finding all steady
states for the above choice of $f$ and second-order coupling.  In this
case, we must solve the system \eqref{eq:eqcond2} depending only on~$s$.
It follows from Fig.~\ref{fig:shapes} that we need to consider only the
regions $\mathcal{I}_{2}$ and $\mathcal{I}_{4}$ with two and four
preimages, respectively.\footnote{Numerically, we will never be exactly at
  an extremum and thus ignore the cases of one or three preimages.}
$\mathcal{I}_{4}$~is non-empty (and an interval), if and only if
$\mu>\mu_{c}\approx1.62377$.  For such values of $\mu$, $\mathcal{I}_{2}$
consists of two intervals with one being infinite; for lower values, it is
one infinite interval.  Practically, we use an auxiliary function
$w_{\mathcal{N}'}(c)$ which for a given value $c\in\mathcal{I}_{M'}$ and an
extended distribution $\mathcal{N}'=(N'_{1},\dots,N'_{M'})$ first computes
numerically all $M'$ solutions $y_{j}$ of $s(y)=c$ and then returns
$c-\gamma\sum_{j}N'_{j}y_{j}$.  Then we compute with a numerical root
finder all zeros of $w_{\mathcal{N}'}$ in $\mathcal{I}_{M'}$.  If no root
exists, then the considered extended distribution does not lead to any
steady state.  Otherwise, each root defines a family of
$\binom{N}{N'_{1}\ N'_{2}\ \cdots\ N'_{M'}}$ steady states.  In our
computations, we also determined for each found family its stability via
the eigenvalues of the Jacobian (all found steady states were hyperbolic).

For finding the real roots of polynomials, i.\,e.\ for determining the
preimages~$y_{j}$, we used the \texttt{Isolate} command from the
\textsc{Maple} \texttt{RootFinding} library which uses for univariate
problems the Approximate Bitstream Newton--Descartes method from
\cite{ksr:real} based on the Rational Univariate Representation from
\cite{fr:rur}.  For finding all zeros of $w_{\mathcal{N}'}$ in a given
interval, we used the \textsc{Maple} \texttt{fsolve} command with the
\texttt{NextZero} option.

As discussed in Section~\ref{sec:css}, our approach to computing all steady
states for a given dimension~$N$ works iteratively starting with $N=1$, as
for any fixed $N$ we determine only those steady states which have no zero
levels; the remaining ones arising from ``lifting'' steady states for lower
dimensions $N'<N$.  For each dimension $N$, we have to consider $C_{N,2}$
extended distributions of length~$2$ and $C_{N,4}$ extended distributions
of length~$4$, thus in total
\begin{displaymath}
  C_{N,2}+C_{N,4}=\binom{N+1}{N}+\binom{N+3}{N}
  =\frac{1}{6}\Bigl(N^{3}+6N^{2}+17N+12\Bigr)
\end{displaymath}
extended distributions.  In the Online Encyclopedia of Integer Sequences
(\url{https://oeis.org/}), this sequence is numbered A003600 and described
as the maximal number of pieces obtained by slicing a torus with $N-1$ cuts
(we have an index shift by one compared to A003600).  Obviously, it grows
cubically with~$N$.

\begin{table}[tb]
  \centering
  \caption{Statistics for the steady states for system dimensions up to
    $N=10$ and for parameter values $\mu=3.5$ and $\gamma=1$.  The first
    row counts the analysed extended distributions, the second row the
    families of new steady states and the third row the total number of new
    steady states.  Similarly, the fourth and the fifth row contain all
    steady states counted in families and in total, respectively.  In the
    sixth row the number of stable three-clade states is given and in the
    last row the required computation time in seconds.}
  \begin{tabular}{l|rrrrrrrrrr}
    $N$ & $1$ & $2$ & $3$ & $4$ & $5$ & $6$ & $7$ & $8$ & $9$ & $10$ \\
    \hline
    ext dist & $6$ & $13$ & $24$ & $40$ & $62$ & $91$ & $128$ & $174$ &
                                                                           $230$ & $297$ \\
    new fam & 4 & 10 & 16 & 24 & 25 & 34 & 40 & 45 & 54 & 67 \\
    new st st & $5$ & $16$ & $52$ & $168$ & $512$ & $2.095$ & $7.254$ & $20.082$
                                                        & $83.030$ & $307.186$ \\
    total fam & 4 & 15 & 31 & 55 & 80 & 114 & 154 & 199 & 253 & 320 \\
    total st st & $5$ & $25$ & $113$ & $489$ & $2.053$ & $8.992$ & $40.736$ &
        $180.599$ & $791.553$ & $3.522.911$ \\
    stable 3 & $0$ & $0$ & $6$ & $24$ & $50$ & $180$ & $602$ & $1.316$ &
        $3.096$ & $10.590$ \\
    \hline
    time & & $6$ & $9$ & $16$ & $33$ & $39$ & $56$ & $73$ & $101$ & $126$ \\
  \end{tabular}
  \label{tab:stat10}
\end{table}

We computed all steady states up to dimension $N=10$ for the parameter
values $\mu=3.5$ and $\gamma=1$.  Table~\ref{tab:stat10} presents some
statistics about the results.  The first row contains the number of
extended distributions that had to be newly studied for a given~$N$,
i.\,e.\ the number $C_{N,2}+C_{N,4}$.  In the second row, the number of
newly found families of steady states is listed, whereas one can find in
the third row the total number of newly found steady states, i.\,e.\ the
sum of the multiplicities of the families.  The fourth and fifth rows count
all steady states, i.\,e.\ including those that arise by ``lifting'' steady
states for lower $N$ by the addition of zero levels, -- again first in
families and then in total.  A numerical fit shows that the growth of the
total number of steady states is exponential; it is well described by the
function $1.29\,e^{1.48N}$.  With $N=10$, we reach already several million
steady states.  This clearly explains why a direct computation of the roots
without exploiting the symmetry will break down very soon (we did not get
further than $N=3$).  As we are particularly interested in stable
three-clade states, their number is recorded in the sixth row.  At the
bottom, the time required for finding new steady states is given in
seconds; these times grow cubically in $N$, i.\,e.\ like the number of
extended distributions to be analysed so that we may conclude that the
computation time per extended distribution is largely independent of~$N$
(one can also observe that the number of families grows only cubically).
The time for ``lifting'' steady states from lower dimensions is negligible.
The total computation required about $7.8$ minutes on a laptop with eight
11th generation Intel i7-11370H cores with a base speed of 3.3 Ghz.  As the
analysis of one extended distribution is completely independent of that of
another extended distribution, the computations can be trivially
parallelised using the \texttt{Grid} package of \textsc{Maple} and the
timings above refer to a parallel computation using all eight available
cores.

\subsection{Bifurcation Analysis for $N=3$}

\begin{figure}[tbp]
  \centering
  \begin{subfigure}[t]{0.45\linewidth}
    \includegraphics[width=\textwidth]{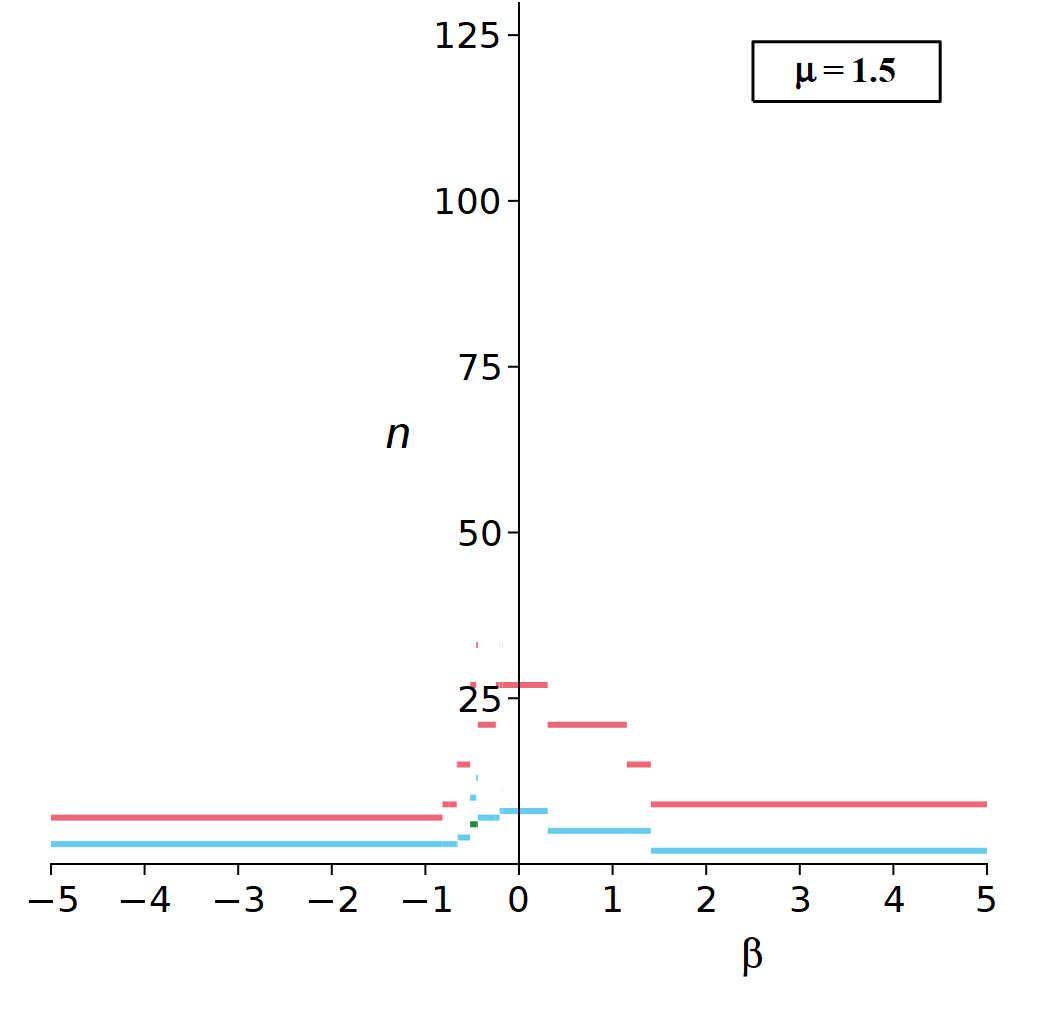}
  \end{subfigure}%
  \hfill
  \begin{subfigure}[t]{0.45\linewidth}
    \includegraphics[width=\textwidth]{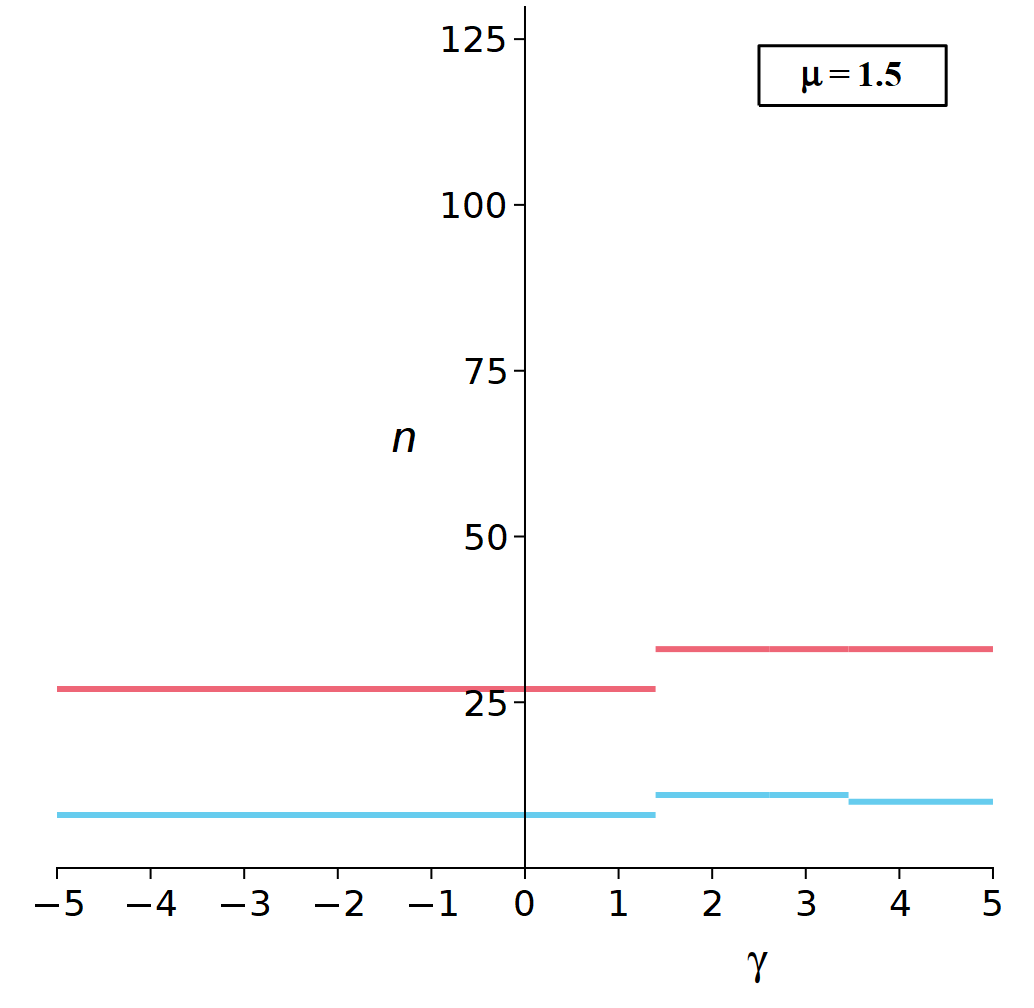}
  \end{subfigure}\\
  \begin{subfigure}[t]{0.45\linewidth}
    \includegraphics[width=\textwidth]{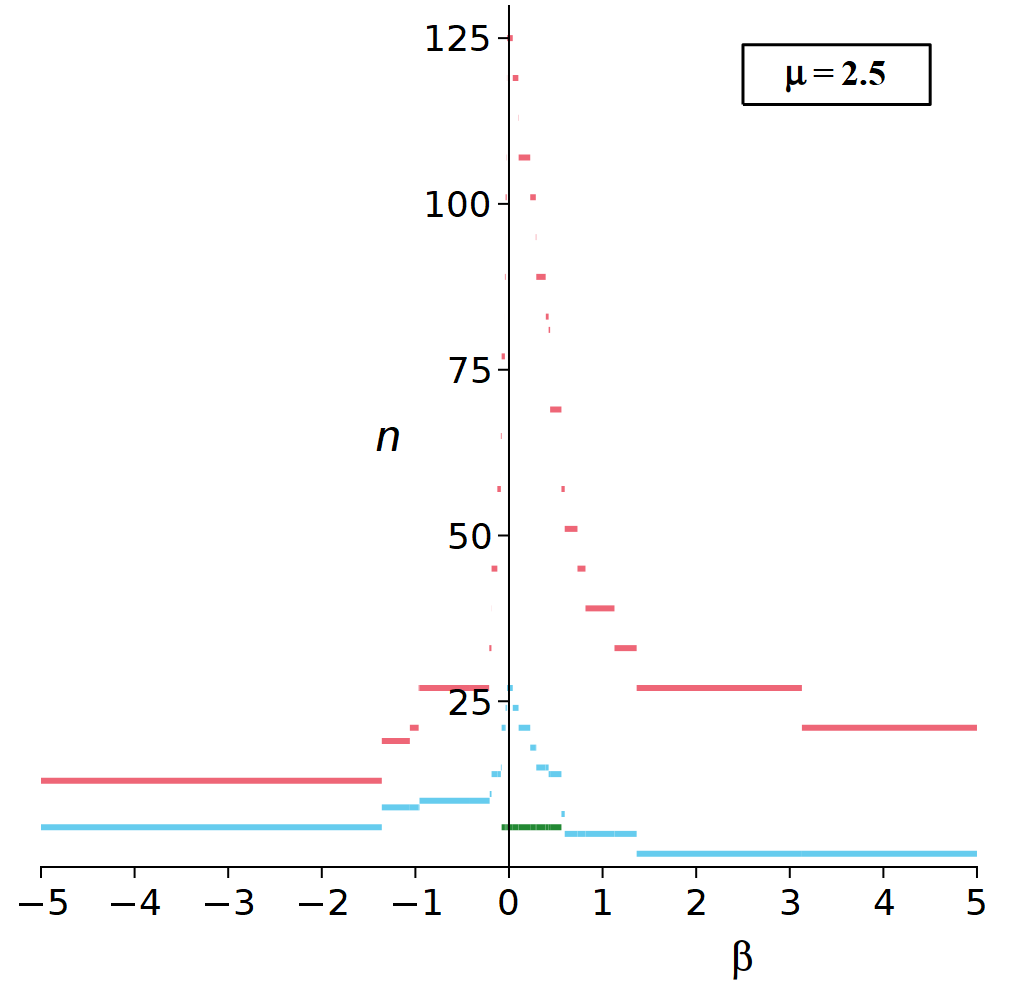}
  \end{subfigure}%
  \hfill
  \begin{subfigure}[t]{0.45\linewidth}
    \includegraphics[width=\textwidth]{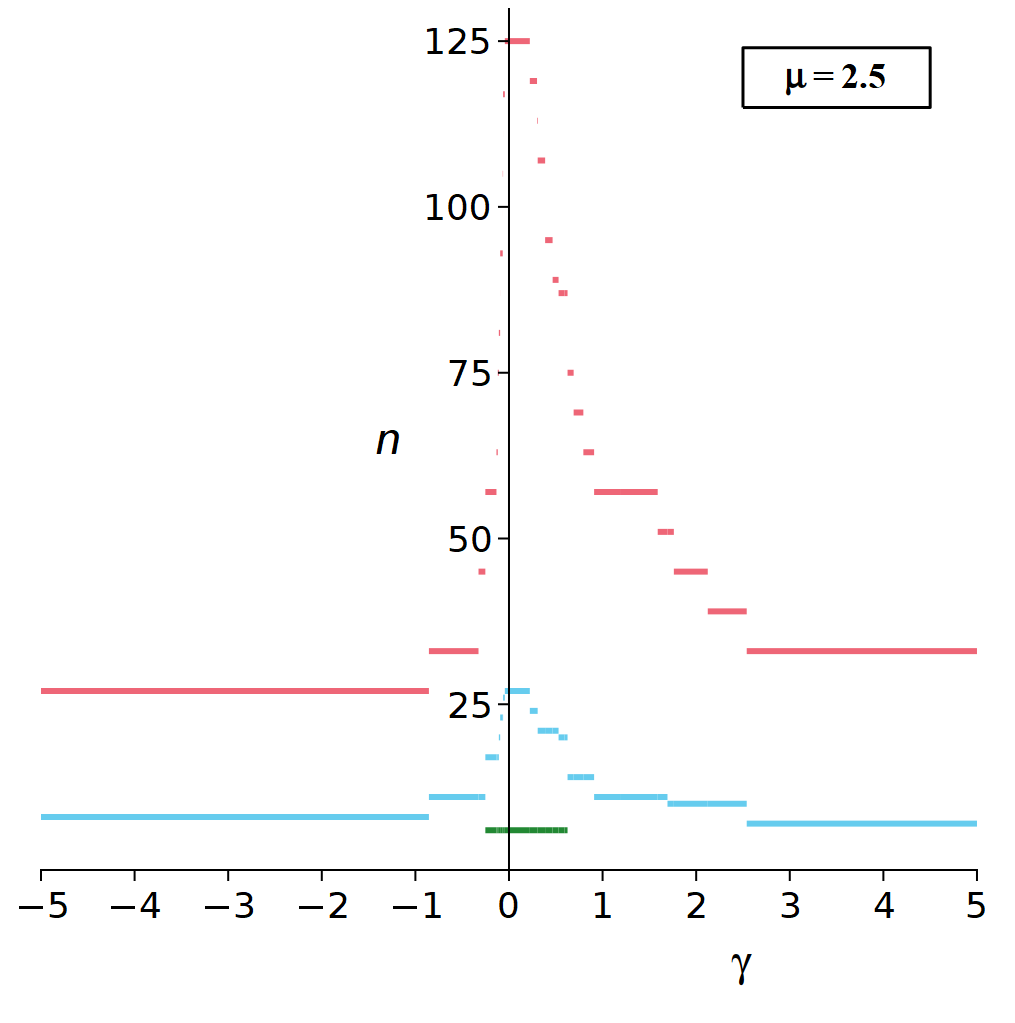}
  \end{subfigure}\\
  \begin{subfigure}[t]{0.45\linewidth}
    \includegraphics[width=\textwidth]{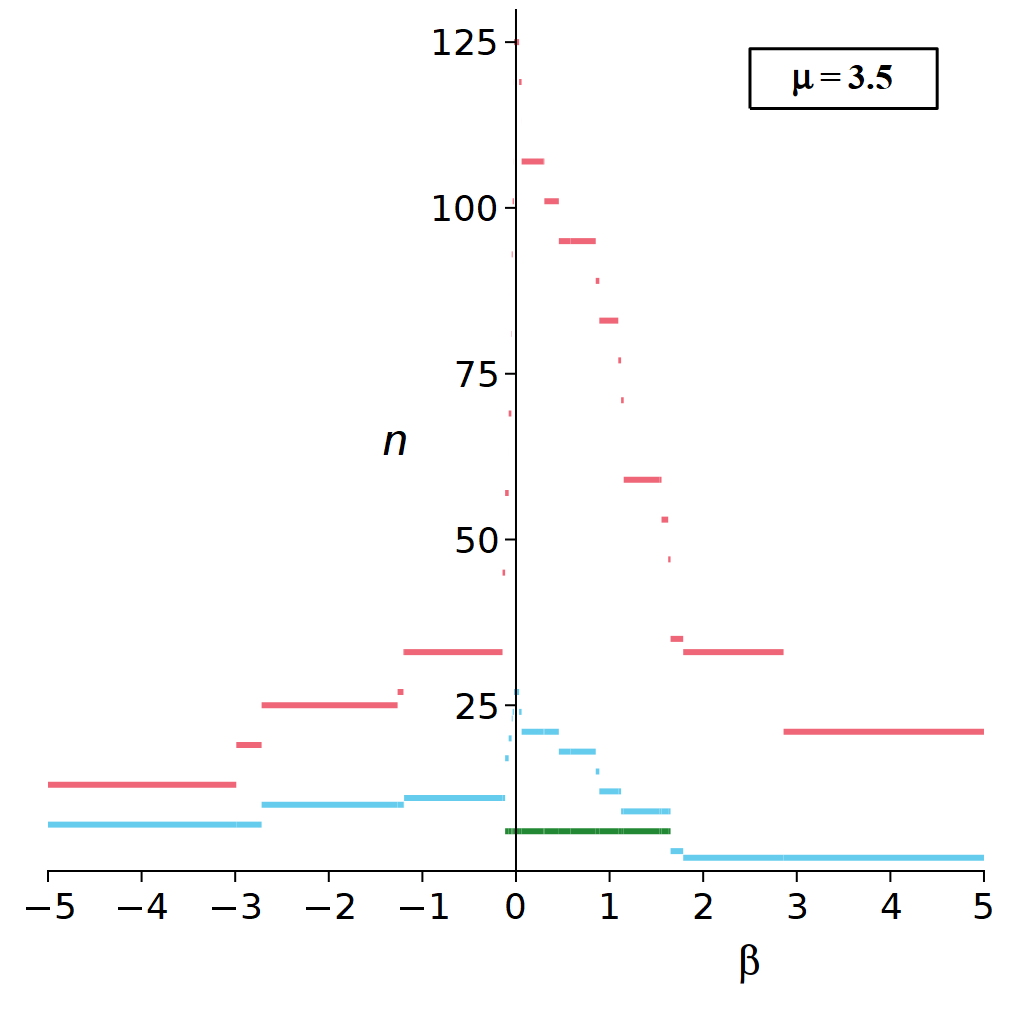}
  \end{subfigure}%
  \hfill
  \begin{subfigure}[t]{0.45\linewidth}
    \includegraphics[width=\textwidth]{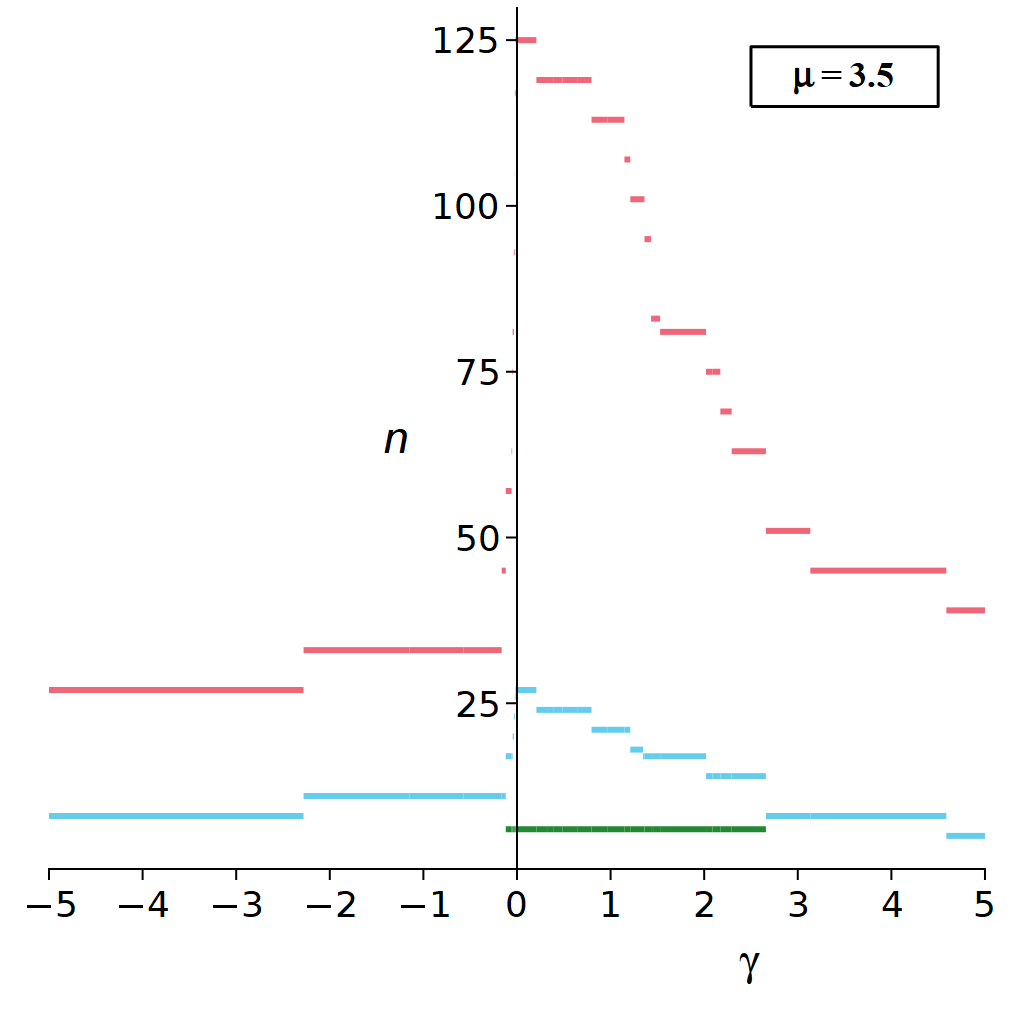}
  \end{subfigure}
  \caption{Statistics of steady states for different parameter values:
    total number (red), stable ones (blue), stable three-clade states
    (green).  The left column is for first-order coupling, the right column
    for second-order coupling.  In the first row $\mu=1.5$, in the second
    one $\mu=2.5$ and in the third one $\mu=3.5$.}
  \label{fig:bifbg}
\end{figure}

For $N=3$ and the polynomial $f$ defined by \eqref{eq:fs}, we numerically
analysed bifurcations of the two model classes described in
Section~\ref{sec:scm} with special emphasis on the appearance of stable
three-clade states.  Of the two parameters -- the coupling strength $\beta$
or $\gamma$, respectively, and the environmental parameter~$\mu$ -- we
always kept one constant and studied the bifurcations arising when the
other one is varied.  For $N=3$, it was still possible to determine
numerically all appearing bifurcation points (typically between $100$ and
$200$).  We used again the \texttt{Isolate} command from the \textsc{Maple}
\texttt{RootFinding} library for determining bifurcation points as the
solution set of a polynomial system. It applies a combination of symbolic
and numeric computations based on the \texttt{RealSolving}
\texttt{C}-library -- see \cite{rz:szds} for the underlying theory.  This
approach guarantees to detect \emph{all} real roots which is important here
for not missing some bifurcation points, as it may be the case with simpler
root finders.

As discussed above, the number of zeros of $f$ changes with $\mu$.  In our
analysis with constant $\mu$, we use the values $1.5$, $2.5$, $3.5$.  For
the first one $f$ has three zeros, for the other two five zeros.  Because
of the $S_{3}$ symmetry, we therefore obtain for $N=3$ without coupling
$3^{N}=27$ and $5^{N}=125$ steady states, respectively; $3!/0!=6$ and
$5!/2!=60$ of these correspond to three-clade states, respectively.  Due to
the permutation symmetry, three-clade states always arise in families of
$3!=6$ for $N=3$; hence we have $1$ or $10$ such families.  All members of
one family obviously show the same stability behaviour.  It follows from
elementary properties of a degree five polynomial of the prescribed form,
i.\,e.\ with a positive leading coefficient, that in the first case at two
of the three zeros and in the second case at three of the five zeros the
derivative is negative.  Hence $2^{N}=8$ and $3^{N}=27$ of the steady
states are stable, respectively, and only in the second case we have one
family of six stable three-clade states.

All bifurcations that we could observe in our numerical experiments concern
changes either in the number of steady states or in their stability.  For
$\mu\neq0,3$, only saddle-node and transcritical bifurcations occur --
sometimes in a degenerate form where both branches have the same stability
or where no swap of stability occurs.  Because of the form of our
polynomial~$s$, at $\mu=0$ and $\mu=3$ highly degenerate bifurcations occur
where four branches intersect.

We discuss first the dependence on the coupling strength $\beta$ or
$\gamma$, respectively.  Fig.~\ref{fig:bifbg} presents some statistics
about the total number of steady states (red), the number of stable steady
states (blue) and the number of three-clade states among these (green).
Independent of the type of coupling, one can make a number of typical
observations.  Firstly, with an increased coupling, the number of steady
states gets typically smaller and becomes constant from a certain value on.
Secondly, for a positive coupling steady states survive much longer than
for a negative coupling.  Thirdly, if the environmental parameter~$\mu$ is
increased (in the figure from the top to the bottom), then the decrease in
the number of steady states becomes slower.  Finally, when the
environmental parameter~$\mu$ has reached a certain threshold, then stable
three-clade states exist mainly only for positive coupling and if the
coupling strength is not too large.  Exceptions from these observations
occur for small values of $\mu$ (less than two), i.\,e.\ in the top row of
Fig.~\ref{fig:bifbg}.  For first-order coupling, stable three-clade states
appear here only for a very small negative $\beta$-interval.  And for
second-order coupling, the number of steady states is increasing with
positive~$\gamma$.

\begin{figure}[tb]
  \centering
  \begin{subfigure}[t]{0.48\linewidth}
    \includegraphics[width=\textwidth]{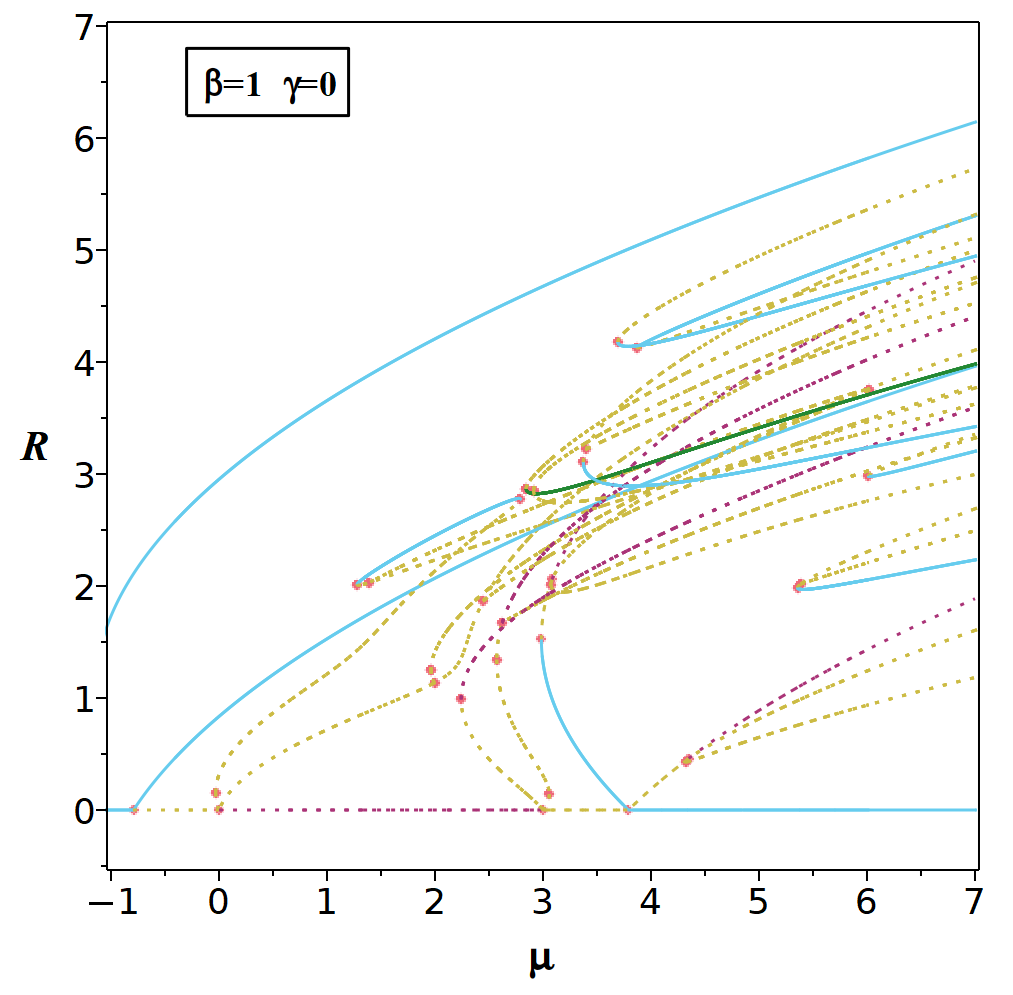}
  \end{subfigure}%
  \hfill
  \begin{subfigure}[t]{0.48\linewidth}
    \includegraphics[width=\textwidth]{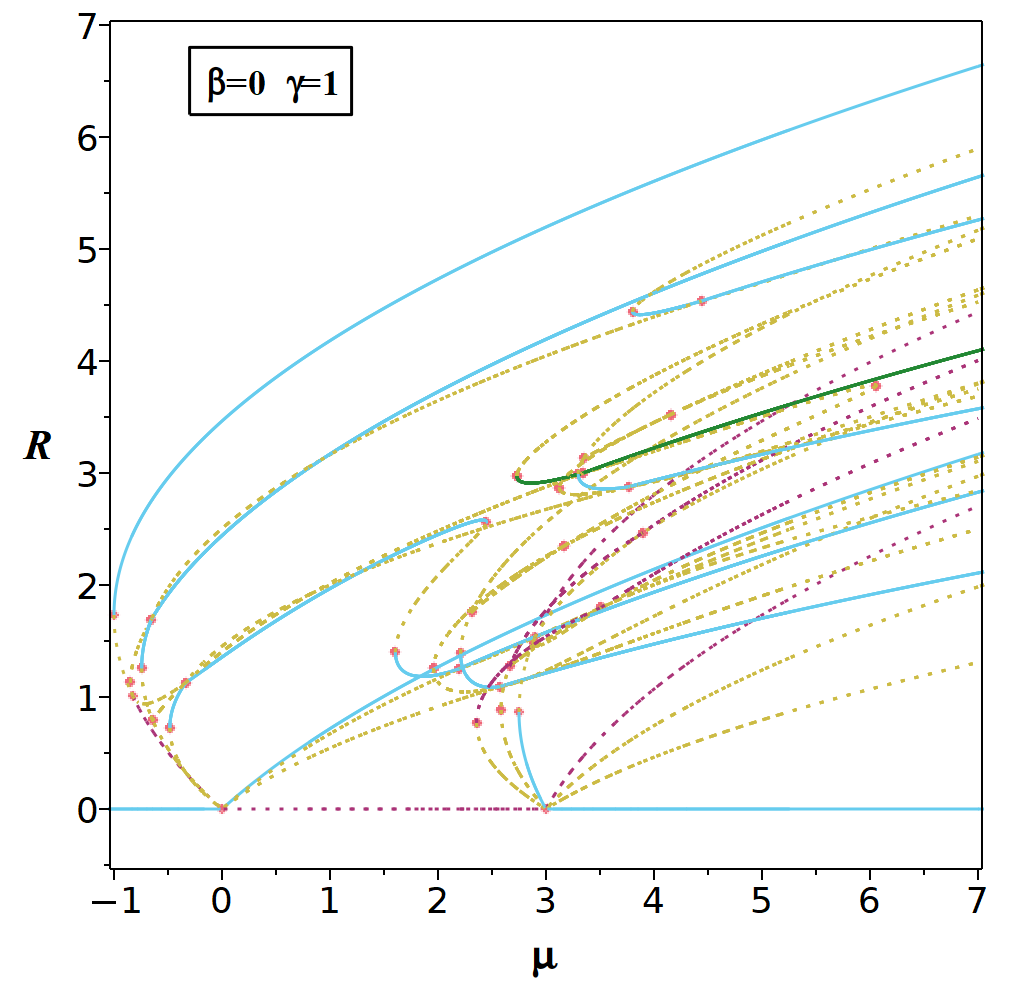}
  \end{subfigure}\\
  \begin{subfigure}[t]{0.48\linewidth}
    \includegraphics[width=\textwidth]{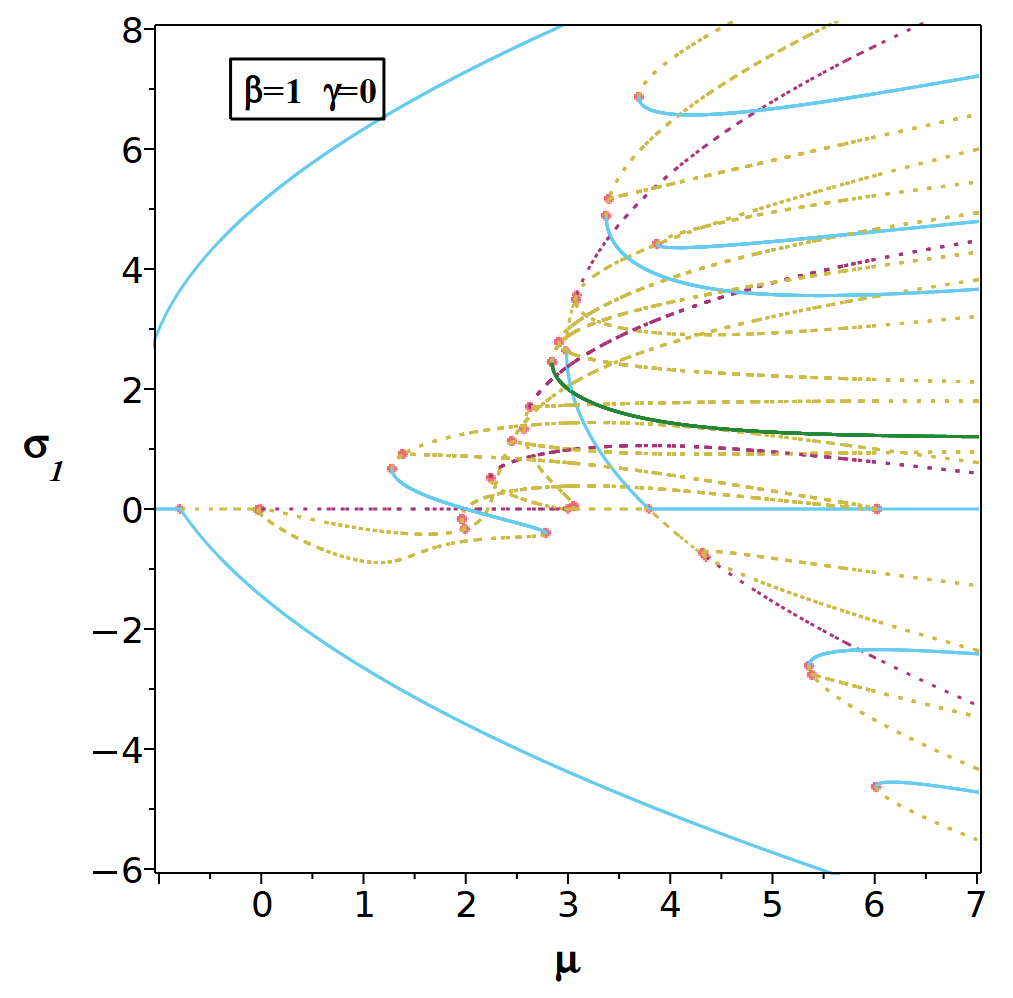}
  \end{subfigure}%
  \hfill
  \begin{subfigure}[t]{0.48\linewidth}
    \includegraphics[width=\textwidth]{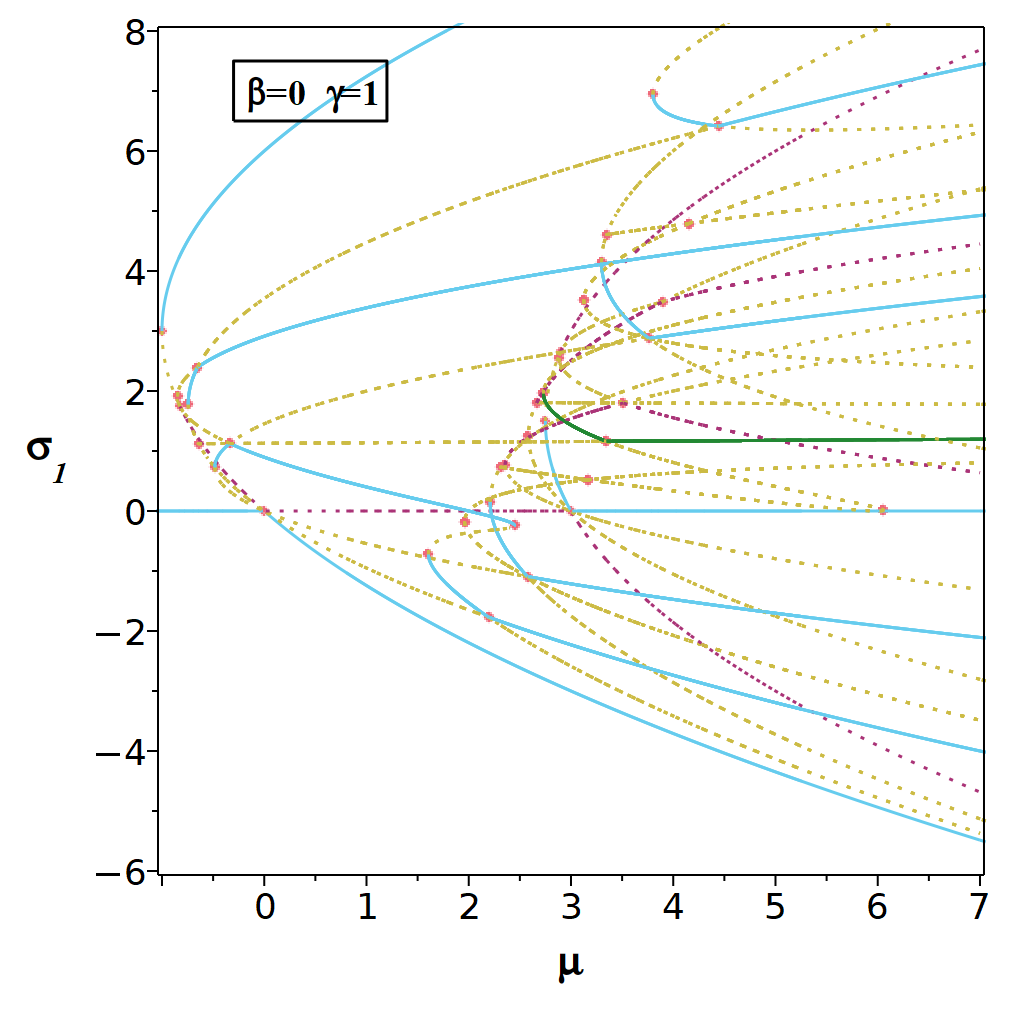}
  \end{subfigure}\\
  \includegraphics[width=0.80\linewidth]{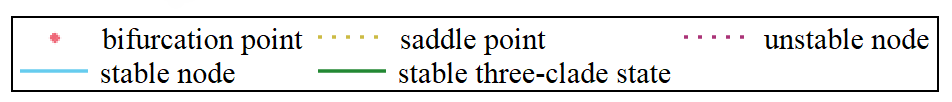}
  \caption{Bifurcation diagrams for varying environmental parameter
    $\mu$. The left column is for first-order coupling with $\beta=1$, the
    right column for second-order coupling with $\gamma=1$. In the top row,
    the distance from the origin is used as ordinate, in the bottom row the
    value of $\sigma_{1}$.}
  \label{fig:bifmu}
\end{figure}

We then proceeded to look for bifurcations, as the environmental
parameter~$\mu$ varies for the specific positive coupling constants
$\beta=1$ and $\gamma=1$, respectively.  We computed numerically
bifurcation diagrams with \textsc{Maple} using a method developed by one of
the authors in \cite{wms:singbif}.  It is based on the Vessiot theory of
general, i.\,e.\ also implicit, differential equations (see
\cite{wms:invol} for a modern introduction and further references) and
consists of deriving a vector field on the $\mu$-$\mathbf{y}$-space such
that the branches are trajectories and the bifurcation points steady
states. By starting trajectory computations at steady states of the
original system for fixed values of $\mu$ between bifurcation points, this
method avoids the need of a complicated analysis of the branching behaviour
even at very degenerate bifurcation points, but currently it can handle
only static bifurcations, i.\,e.\ bifurcations involving only steady states
and not more complicated structures like limit cycles.

We found $98$ and $141$, respectively, bifurcation points lying in
$[-1.2,14.5]$ and $[-1,46.7]$, respectively.  These high numbers are again
a consequence of the symmetry, as the determining system for the
bifurcation points inherits it.  Fig.~\ref{fig:bifmu} shows the
corresponding bifurcation diagrams. As ordinate we use in the top row the
distance $R=\|\mathbf{y}\|$ of the steady state to the origin and in the
bottom row the value of $\sigma_{1}$ (here in particular the sign is of
interest).  A colour coding has been used for the branches: green depicts
those stable nodes that are three-clade states and blue the remaining
stable nodes; red marks unstable nodes and yellow saddle points.  In
addition, stable branches are plotted in solid style and unstable branches
in dotted style.  Note that as a consequence of using $S_{N}$-invariant
quantities as ordinates, all steady states related by a permutation
coincide in the diagrams.  Hence they show much less bifurcation points and
branches as there actually exist which allows for a better interpretation.

Most of the bifurcation points lie in the shown range $-1\leq\mu\leq7$.  In
both cases, most bifurcations are of saddle-node type (sometimes of a
degenerate form with two branches of the same stability).  In addition,
some transcritical bifurcations appear (again sometimes of a degenerate
form without an exchange of stability).  Only for the two special values
$\mu=0$ and $\mu=3$ highly degenerate bifurcations appear where eight
branches meet.  The diagram confirms the statistics from
Fig.~\ref{fig:bifbg} that there are much more unstable steady states than
stable ones; in particular, saddle points abound.

\subsection{(In)Accessibility of Stable Three-Clade States for First-Order
  Coupling and $N=3$}

In numerical simulations starting at a small random perturbation of the
origin, one can observe that for some parameter values it is very easy to
end in a stable three-clade state while for other values this is very rare
(see the more detailed discussion in later sections).  The location of the
green branch in Fig.~\ref{fig:bifmu} indicates one possible reason: other
stable steady states corresponding to only one or two clades exist closer
to the origin which may attract trajectories.  Furthermore, there are many
saddle points around; in fact, for larger $N$ the vast majority of steady
states are saddle points.  Relevant in this context are saddle points where
one invariant manifold is a hypersurface, i.\,e.\ of dimension $N-1$.  Such
invariant manifolds act as separatrices, as trajectories cannot cross them;
therefore we call them \emph{separating saddle points}.\footnote{For $N=3$,
  obviously every saddle point is separating; for larger $N$, they are
  getting rarer and rarer.  In the above discussed computations of all
  steady states for $\mu=3.5$ and $\gamma=1$, i.\,e.\ for a second-order
  coupling, we also determined the type of each steady state and for saddle
  points in addition whether they are separating.  For $N=3$, $74\%$ of the
  steady states are separating saddle points, for $N=4$ still $50\%$, for
  $N=6$ only $20\%$, for $N=8$ $7.5\%$ and finally for $N=10$ the
  percentage drops to $2.6\%$ (which, however, actually means that close to
  $100.000$ separating saddle points exist).}  Depending on their location,
they can effectively shield off a stable three-clade state from the origin.
However, as in Fig.~\ref{fig:bifmu} the ordinate only shows the distance
from the origin, these observations are only indications not allowing for a
clear conclusion.

We therefore numerically studied the (in)accessability of the stable
three-clade states for $N=3$ and first-order coupling with $\beta=-0.3$ in
the following manner: we chose points on the surface of a small sphere
around the origin and integrated the trajectories through them until they
reached a stable node; then we marked the initial point in a colour
corresponding to this stable node.  To ensure that we correctly sample the
surface of the sphere, we used a Fibonacci lattice for the initial points,
as for it the Voronoi cell of each point has approximately the same area.
Thus a simple counting of the points leading to a stable three-clade state
(without any weights) yields a good approximation of the probability that a
random perturbation ends in such a state \citep{ag:fib,sp:fib}.

\begin{figure}[tb]
  \centering
  \begin{subfigure}[t]{0.45\textwidth}
    \includegraphics[width=\textwidth]{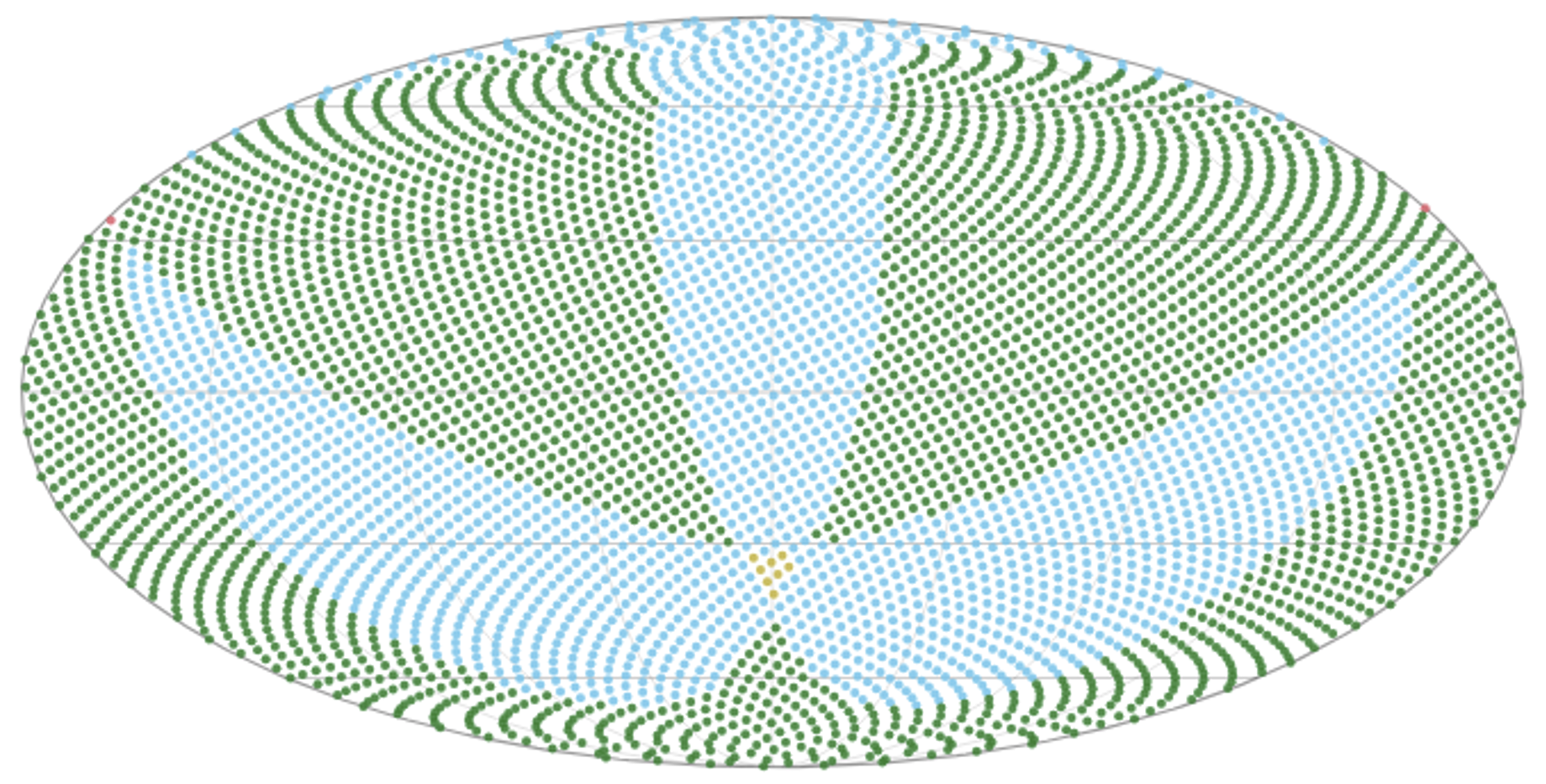}
    \caption{Mollweide projection of the initial points on a sphere of
      radius $R=0.05$ for $\mu=1.8$.}
  \end{subfigure}%
  \hfill
  \begin{subfigure}[t]{0.45\textwidth}
    \includegraphics[width=\textwidth]{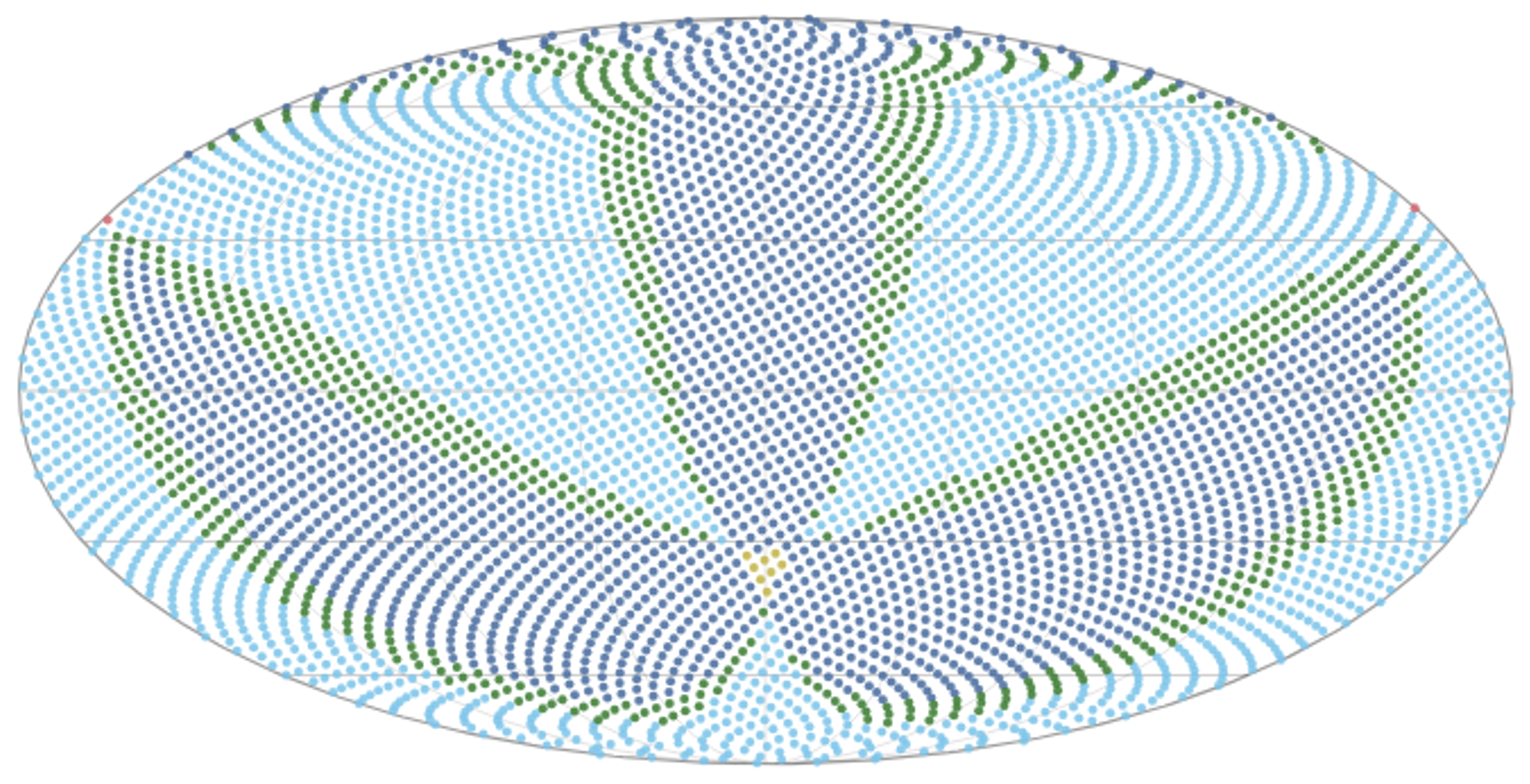}
    \caption{Mollweide projection of the initial points on a sphere of
      radius $R=0.05$ for $\mu=1.85$.}
  \end{subfigure}\\
  \begin{subfigure}[t]{0.45\textwidth}
    \includegraphics[width=\textwidth]{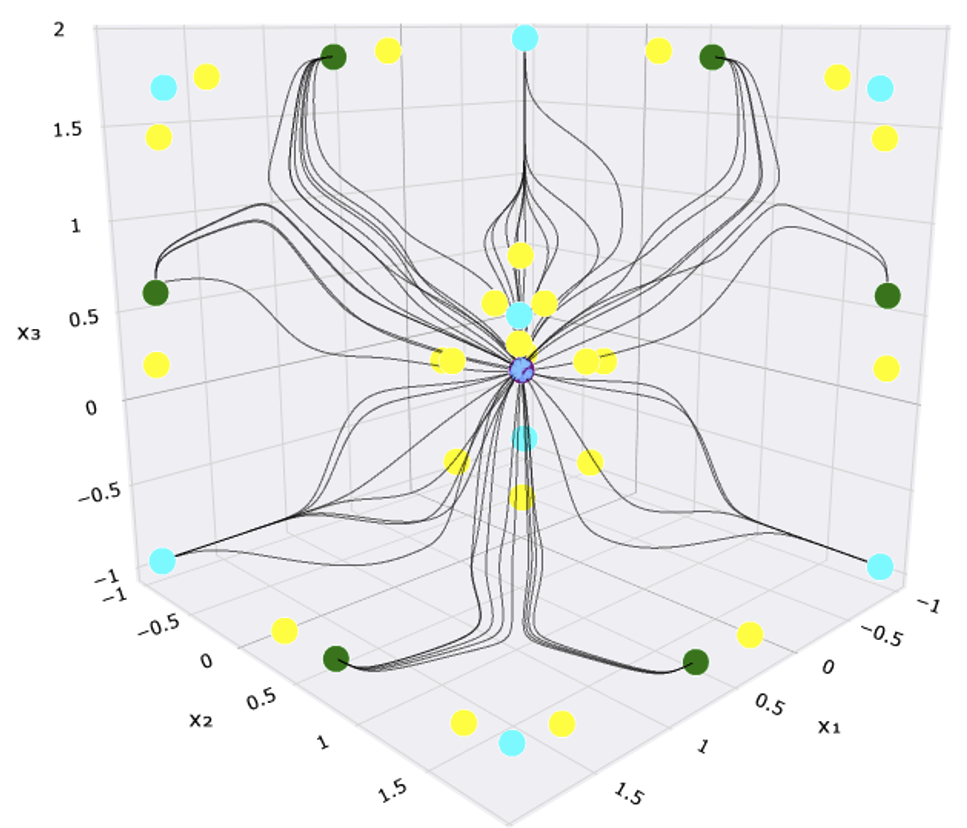}
    \caption{Phase portrait for the same parameter values as above.}
  \end{subfigure}%
  \hfill
  \begin{subfigure}[t]{0.45\textwidth}
     \includegraphics[width=\textwidth]{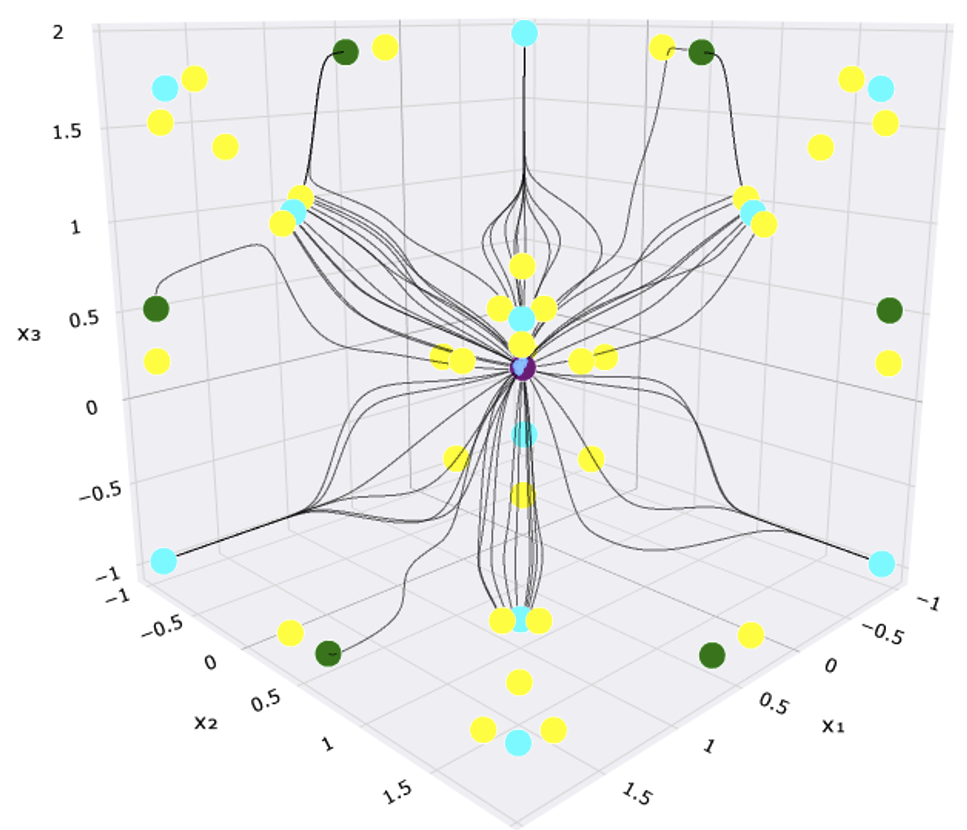}
    \caption{Phase portrait for the same parameter values as above.}
  \end{subfigure}\\
  \caption{Top row: the basins of attraction for different stable nodes on
    the surface of a small sphere around the origin.  Green points
    represent initial data ending in a three-clade state; the blue and cyan
    points represent points ending in two-clade states, red and yellow
    indicate one-clade attractors.  Bottom row: some trajectories starting
    at random points on the sphere.  Here again the green points represent
    the family of stable three-clade states and the cyan points other
    stable nodes; the saddle points are indicated in yellow, the (unstable)
    origin in purple. The radius of the sphere was $R=0.05$ and we used
    first-order coupling with $\beta=-0.3$.  Left column: $\mu=1.8$. Right
    column: $\mu=1.85$.}
  \label{fig:moll}
\end{figure}

For representing the results graphically, we used the Mollweide projection
to project the surface of the sphere to the plane.  It is an equal-area
projection \citep{jps:mapproj} and thus provides a proper impression of the
relative frequency of the different cases, although the attractors are
distorted, in particular, close to the poles.  In this pseudo-cylindrical
projection, the surface is mapped into an ellipse where the horizontal
semiaxis has twice the length of the vertical one.  For better orientation,
the plots in Fig.~\ref{fig:moll} show in grey some longitudes (curved
verticals) and latitudes (horizontal lines).  To make the symmetry of the
plots more clearly visible, the plots were turned by $135^{\circ}$ so that
the middle axis corresponds to this angle and not to $0^{\circ}$.  The
symmetry simply stems from the fact that the 3D diagonal contains all
one-clade states and all two clade-states lie symmetric to it in the
corresponding plane $x_{k}=x_{\ell}$.

Fig.~\ref{fig:moll} contains in the top row the results for a first-order
coupling with $\beta=-0.3$ and two different values of $\mu$, namely
$\mu=1.8$ and $\mu=1.85$.  The $5.000$ initial points were always on a
sphere of radius $R=0.05$.  We always observed as attractors five
different families of stable nodes: one family of three-clade states shown
in green, two families of two-clade states shown in blue and cyan,
respectively, and two one-clade states shown in yellow and red,
respectively.

It turned out that although the two used values of $\mu$ are rather close,
one obtains very different results.  For $\mu=1.8$ about $61\%$ of the
computed trajectories ended in a three-clade state, for $\mu=1.85$ only
about $16\%$.\footnote{If one increases $\mu$ further, the percentage
  rapidly drops.  At $\mu=1.9$, it is only about $0.3\%$ and for still
  larger values, we could not observe any trajectories leading to
  three-clade states.}  The one-clade states were rather rare: yellow
points appear a few times in an ``triangle'' a bit south of the centre of
the plots and in each plot there are only two red dots on the boundary at a
bit more than $30^{\circ}$ north.  For $\mu=1.8$, only one family of
two-clade states appears as attractors; for $\mu=1.85$, both families
appear frequently and their basins of attraction are separated by the
remaining initial positions leading to a three-clade state.

The phase portraits shown in the bottom row of Fig.~\ref{fig:moll} offer a
partial explanation for these observations.  For $\mu=1.8$, only few other
stable steady states lie closer to the origin than the three-clade states
and hence a larger percentage of the trajectories reach the latter ones.
For $\mu=1.85$, one can find more other stable steady states in the
vicinity of the origin and these are obviously fairly successful in
attracting trajectories starting close to the origin.  If $\mu$ is
increased further, this effect becomes more and more pronounced, until
other stable steady states form a kind of ``cage'' around the origin
catching all trajectories emerging from a neighbourhood of the origin.  It
is notable that most trajectories show sharp bends.  We suspect that these
are due to the separatrices emerging from the saddle points, but this
conjecture must be further analysed.

\subsection{Persistence of Stable Three-Clade States for Second-Order
  Coupling and $N=5$}
\label{sec:N5} 

In a further extensive computation, we studied over a larger range of
values of the environmental parameter~$\mu$ how the stable three-clade
states evolve for $N=5$ and second-order coupling. Since we also took all
zero levels into account, this required to analyse a larger number of
(extended) distributions.  The starting point are the possible
\emph{occupation patterns}.  We write occupation patterns for three-clade
states in the form $\{N_{1},N_{2},N_{3}\}$ with $N_{1}+N_{2}+N_{3}=N=5$ and
$N_{1}\geq N_{2}\geq N_{3}$.  Such a pattern comprises all (extended)
distributions where for each $N_{i}$ one level exists that appears $N_{i}$
times as a coordinate of the steady state (note that, by contrast to a
distribution $(N_{1},N_{2},N_{3})$, it is not necessarily the lowest level
that appears $N_{1}$ times).  As zero levels require a special treatment,
we indicate them by an index~$0$.

The following six occupation patterns are possible for $N=5$: $(3,1,1)$,
$(2,2,1)$, $(3,1,1_{0})$, $(2,2,1_{0})$, $(2,2_{0},1)$, $(3_{0},1,1)$.  For
each pattern, there exists a certain number of ways how it is realised as a
three-clade extended distribution.  By an elementary combinatorial
computation,\footnote{These numbers arise as multiset permutations: we
  always have four slots and taking the zero occupation into account we
  have in four patterns the situation that there are three numbers, one of
  them double, and in two patterns there are only two numbers, both double.
  The sum of the numbers is always four.  Thus we get the multinomial
  coefficients $\binom{4}{2\ 1\ 1}=12$ and $\binom{4}{2\ 2}=6$.} these are
$12$ for all patterns except the last but first and the last but third
where only $6$ possibilities exist.  In total, for $60$ extended $4$-fold
distributions we had to compute a diagram like the one shown in
Fig.~\ref{fig:ccrit} plus a diagram showing the evolution of the three
levels.  For this purpose, we computed the critical $c$-value (if one
existed) plus the corresponding levels on a discrete grid of $\mu$-values
and then determined for each found steady state its stability.

\begin{figure}[ht]
  \centering
  \includegraphics[width=9cm]{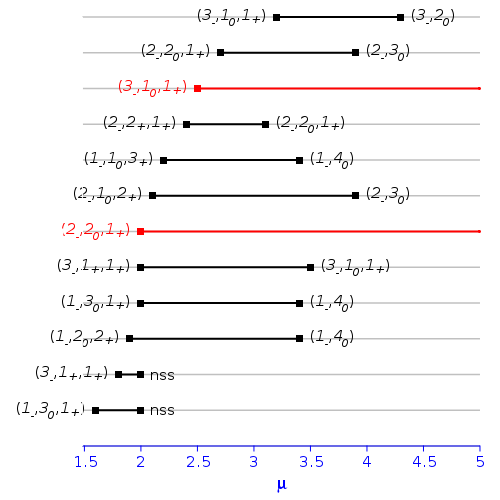}
  \caption{Persistence of stable three-clade states for $N=5$, $\beta=0$
    and $\gamma=-1$ (including states with zero levels).  To the left of
    each bar, the state is described by a three-clade distribution; to the
    right, its ``fate'' is indicated: \texttt{nss} means the steady state
    disappears, otherwise the state into which it evolves is given.  Only
    the two states in red persist over the whole considered range of
    $\mu$.}
  \label{fig:persist3splits}
\end{figure}

For a second-order couping with $\gamma=-1$, only $12$ among these $60$
extended $4$-fold distributions lead for at least some range of
$\mu$-values to stable steady states.  Fig.~\ref{fig:persist3splits} shows
the corresponding non-extended distributions (in two cases, the
non-extended distributions were equal) together with their ranges of
existence.  An index at the occupation numbers indicates whether the
corresponding level is negative, positive or zero.  In two cases, the
steady state simply disappeared at the end of the range (denoted in
Fig.~\ref{fig:persist3splits} by \texttt{nss}); if one plots here a diagram
like in Fig.~\ref{fig:ccrit}, then the red curves hits after some time
again one of the blue boundaries and thus stops.  In most cases, one of the
non-zero levels became zero at a certain $\mu$-value (and for larger
$\mu$-values the steady state was then unstable).  As a zero level can
never be left, this means that we either moved to another case with a
higher occupation of the zero level or that the three-clade state evolved
into a two-clade state.

It is interesting to note that several three-clade states evolved at
exactly the same $\mu$-value into exactly the same two-clade state.  This
is not evident from Fig.~\ref{fig:persist3splits}, as there the level
values are not indicated.  But our numerical computations show that in each
case exactly the same levels occur.  At least in the case where three
different states merge into one, this may be considered as a degenerate
bifurcation. 

Only two possibilities lead to a persistently stable three-clade state;
each of them involving a zero level.  Around $\mu=2$, the steady state
corresponding to the extended $2$-fold $N=3$ distribution $(2,0,0,1)$
augmented by a doubly occupied zero level appeared as a stable state and
persisted until the end of our computation for $\mu=7$ (and the form of the
diagrams indicates that it will probably persist for a long time after).
Around $\mu=2.5$, in addition the steady state corresponding to the
extended $2$-fold $N=4$ distribution $(3,0,0,1)$ augmented by a singly
occupied zero level appeared as a stable state and also persisted until the
end of our computation.

\section{Trajectory Simulations}\label{sec:trajsim}

In this section we complement the numerical stability and accessibility analysis by simulations of trajectories in order to find the predicted three-clade states. All trajectories were simulated via self-written Python programs using SciPy routines \texttt{odeint()} and \texttt{solve\_ivp()} for numerical integration of ordinary differential equations. We limit ourselves to a thorough analysis of $N\le5$ and give examples for larger $N$ at the end of the section.
\subsection{Attractor States for $N \le 5$}
As strategy for finding the location of attracting states, we start from all (stable and unstable)  roots of the uncoupled system and follow the trajectories for different values of the coupling parameter. Since the starting values are simply given by the roots of the polynomial $f(x)$ they are easily calculated.
We expect that, at least for weak coupling, the shifted stable states of the uncoupled system remain attractors. For accounting we look for all combinations of the up to three stable levels for a single population in the network of $N$ populations. 
Recall that the total number of states disregarding degeneracy (i.e. permutation of populations without changing the values and occupations of the levels) is given by the combination with repetition as 
\begin{equation}
  \label{eq:comb_with_rep}
  {C}_{N,M'} = \binom{N+M'-1}{N} 
\end{equation}
in which $N$ is the number of populations and $M'$ the number of levels (here 3, for the quintic polynomial).

For example, if we had three populations ($N=3$), we expect 
\begin{equation}
  \label{eq:K_32}
  {C}_{N,M'} = \binom{3+3-1}{3}=\frac{5!}{3!2!}=10
\end{equation}
different extended distributions (families of states) that can be represented as $(1,1,1)$, $(2,1,0)$, $(2,0,1)$, $(1,2,0)$, $(0,2,1)$, $(1,0,2)$, $(0,1,2)$, $(3,0,0)$, $(0,3,0)$, and $(0,0,3)$. 

Thus we expect one three-clade distribution, six different two-clade distributions (two populations form one clade, the third one the second), and three different distributions in which all populations have the same trait. In the notation, however, we do not distinguish between e.g. $(2,1,0)$, $(2,0,1)$ and $(0,2,1)$ and denote all as $(2,1)$ for simplicity. Since the occupation numbers are sorted with increasing values of $x$, $(2,1)$ and $(1,2)$, belonging to the same pattern, are kept distinct. 

For $N=2$, there is no three-clade state possible, for $N=4$ we have three three-clade states: $(2,1,1)$, $(1,2,2)$, and $(1,1,2)$, and for $N=5$ we have six three-clade states: $(3,1,1)$, $(1,3,1)$, $(1,1,3)$, $(2,2,1)$, $(2,1,2)$, and $(1,1,2)$.
Since in the simulations we do not treat zero values specially, we do not distinguish here $(2,2,1)$ from $(2,2_0,1)$ in the case of second order coupling. States like the latter are metastable and lead to trajectories that stay pinned to zero. We avoid such a situation by a small non-zero value for $r$ here, or (for $r=0$) we introduce a fluctuation of the starting conditions. We cannot avoid, however, that trajectories with $\dot x$ values near to zero slow down significantly.

Of all these possible states, some loose stability upon a change in the bifurcation parameter or increasing the coupling as it was discussed in the previous section. Table ~\ref{tab:OV} gives an overview from simulations for $\mu=2.5, r=0.001$ but varying $N$ how the number of stable distributions decreases when the coupling parameter (first or second order) increases. 

\begin{table}[tb]
    \centering
    \caption{Number of accessible distribution families for different coupling models.  $\mu=2.5$, $r=0.001$, other parameters are as before.}
    \begin{tabular}{llllll}
         \hline
         \textbf{} && \textbf{N=2} &  \textbf{N=3} & \textbf{N=4} & \textbf{N=5}\\
         \hline
         uncoupled && 6 & 10 & 15 & 21 \\
         $\beta=+0.1/N,\gamma=0$ &&  6 &  10 & 15  & 20 \\
         $\beta=-0.1/N,\gamma=0$ && 5  &  9 & 13  &  17\\
         $\beta=+1,\gamma=0$ &&  3 &  3 &  3 & 3 \\
         $\beta=-1,\gamma=0$ && 3  &  5 & 5  & 8 \\
         $\gamma=+0.1/N,\beta=0$ && 6  & 10  & 15  & 21 \\
         $\gamma=-0.1/N,\beta=0$ && 6  &  10 &  15 &  21\\
         $\gamma=+1,\beta=0$ && 4  & 5 & 8  & 8 \\
         $\gamma=-1,\beta=0$ && 3  & 4  & 6  & 8 \\
          \hline
    \end{tabular}
    \label{tab:OV}
\end{table}

\begin{figure}[tbp]
    \centering
    \includegraphics[width=0.9\textwidth]{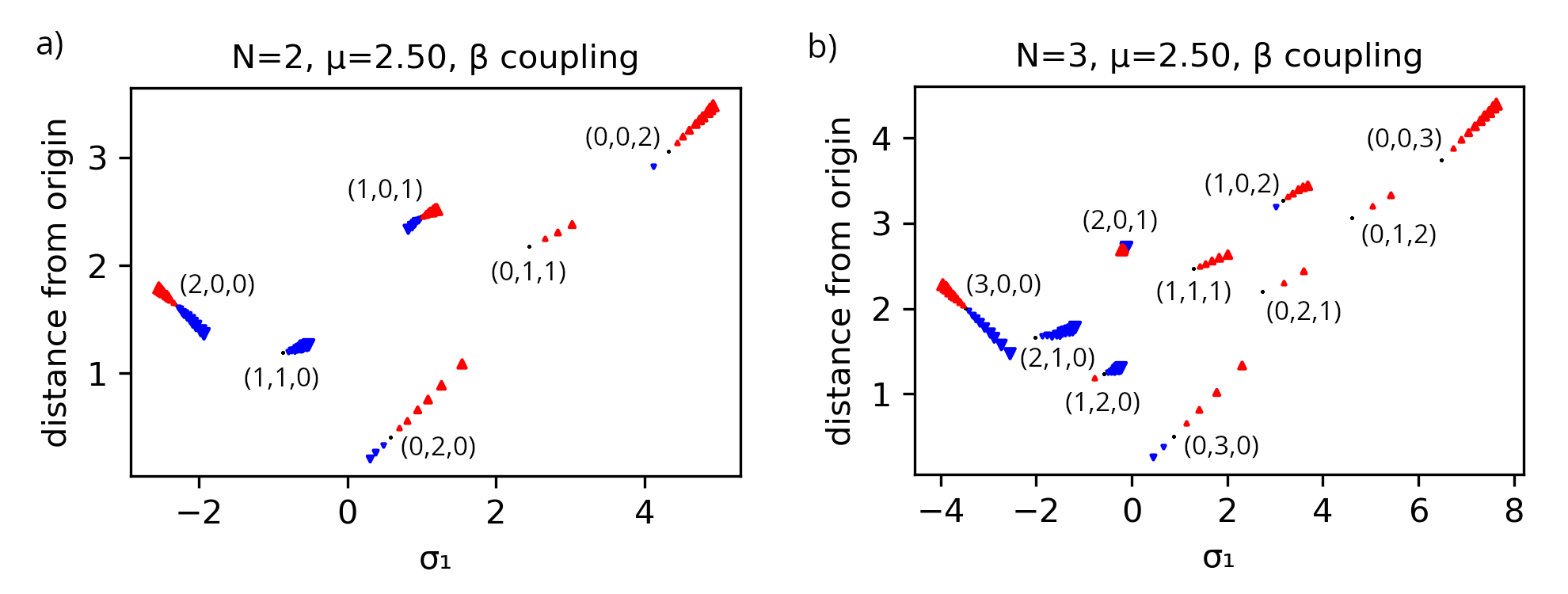}
    \caption{Shift of attracting steady states for a) $N=2$ and b) $N=3$ for fixed $\mu=2.5, r=0$, and first order coupling. To represent the state $\mathbf x$, the length of the vector was plotted against the sum of the components ($\sigma_1$). The points show endpoints for trajectories starting at the steady states of the uncoupled system with a random fluctuation of maximal amplitude ($\pm0.0025$). Positive values of the coupling are indicated as red triangles upwards, negative values as blue triangles downwards. Extended distributions are assigned according to the occupied levels.}
    \label{fig:statesN23}
\end{figure}

The shift of states within the range of stability determined by the coupling parameter can be visualized for any dimension by plotting state diagrams in which the states  are represented by the length of the vector $\mathbf x$ and the sum of its components, i.e. the elementary polynomial $\sigma_1$. Both quantities are invariant against permutations of the components, so the family of states belonging to an extended distribution is characterized by a single point in the plot (accidental overlaps of distributions may occur, though). In Fig.~\ref{fig:statesN23}, the location of the 6 possible extended distributions for $N=2$ and the 10 extended distributions for $N=3$ is shown in the case of first order coupling.
One can see the location and stability limits for the different states that have been summed up in Fig. \ref{fig:bifbg}. For example, the $N=2$ distribution (1,1,0) is not stable for positive values of $\beta$ and the trajectory is redirected to (2,0,0). Likewise, for higher positive values, (0,1,1) and the (0,2,0) end in the extended distribution (0,0,2). So the number of states is continuously reduced.

\begin{figure}[tbp]
    \centering
    \includegraphics[width=0.9\textwidth]{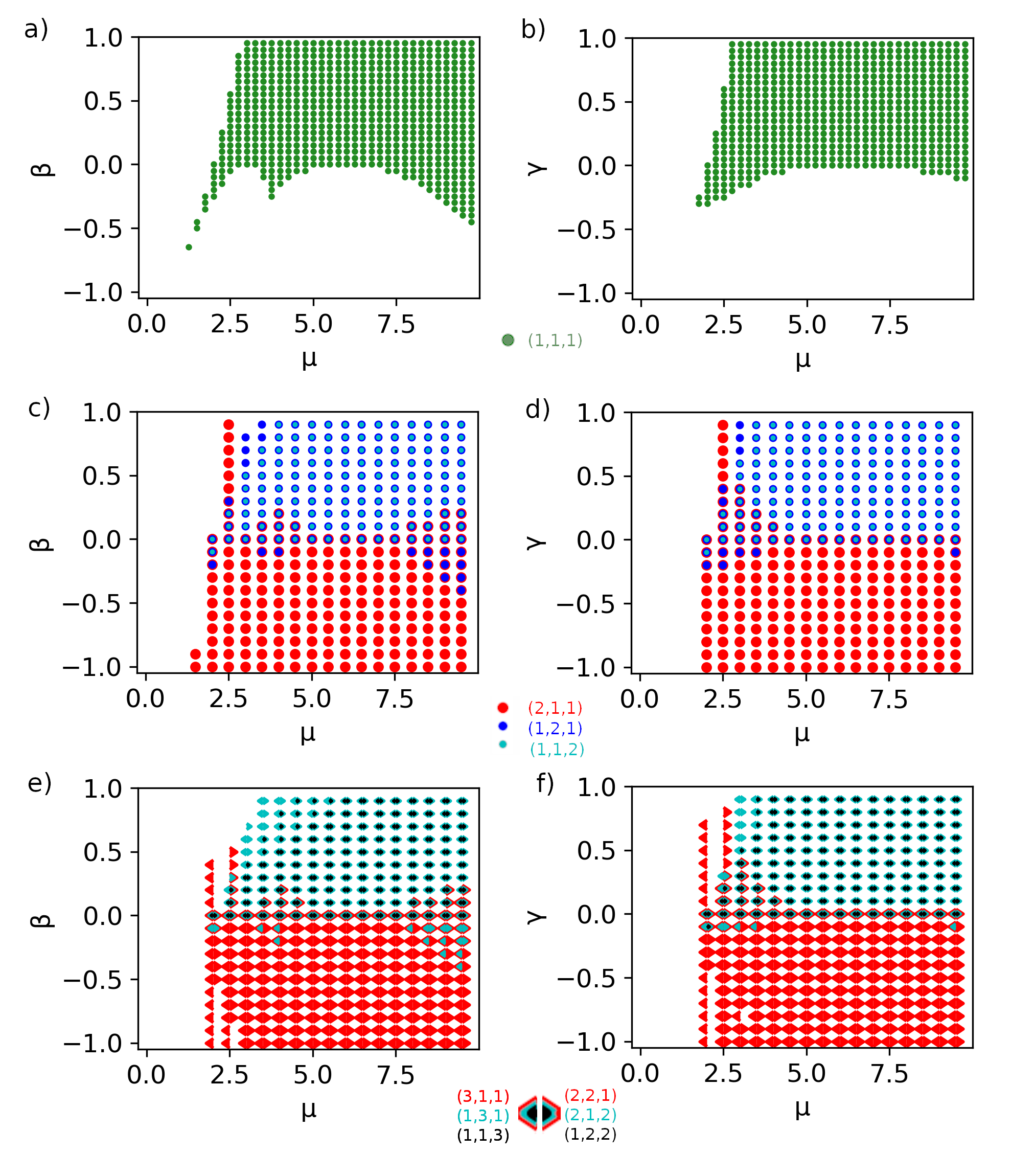}
    \caption{Occurrence of the three-clade distributions for $N=3,4,5$ and different values of $\mu$ and the coupling parameters $\beta$ or $\gamma$, the other parameter set to zero, respectively. All parameters as in fig. \ref{fig:bifmu} ($r=0$). Starting conditions are the uncoupled states with a fluctuation of $\pm0.0025$. The size and colour coding of the distributions is explained in the corresponding legends. For $N=3$ (a and b), the $(1,1,1)$ distribution is displayed in green. For $N=4$ (c and d), the three possible distributions are represented as circles with decreasing size. For $N=5$ (e and f), the distributions $(3,1,1)$, $(1,3,1)$, and $(1,1,3)$ are represented by triangles to the left in decreasing size, and the distributions $(2,2,1)$, $(2,1,2)$ and $(1,2,2)$ as triangles to the right, also in decreasing size.}
    \label{fig:3splits}
\end{figure}

With $N=3$, we meet the first case of a three-clade distribution which is unique
and denoted as $(1,1,1)$. Also this distribution
has a limited stability range, as can be seen in Fig.~\ref{fig:statesN23}b for small positive values of $\beta$). Since in these diagrams either the coupling parameter or $\mu$ can be varied, we abandon now the information about $\mathbf{x}$ and mark just the occurrence of three-clade distributions in $\mu-\beta$ and $\mu-\gamma$ diagrams. The total stability range of the (1,1,1) distribution for different $\mu$ values and coupling constants with the same parameters as in Fig.~\ref{fig:statesN23} is displayed in Fig.~\ref{fig:3splits}a). We confirm the finding from the bifurcation analysis in  Fig.~\ref{fig:bifbg} that three-clade states are confined mainly to positive coupling with some details on the stability range. As example, for $\mu=2.5$ in the case of first order coupling, the stability is limited to the interval $\beta\in[0,+0.55]$ (cf. also Fig.~\ref{fig:statesN23}b and Fig.~\ref{fig:bifbg}). If $\beta$ is smaller, trajectories are
attracted to $(2,0,1)$, if it is larger to $(1,0,2)$. 
The stability diagram for second order coupling (~\ref{fig:3splits}b) is similar but less fine-structured, with occurrence mainly for positive coupling.
For both coupling schemes, a singular situation theoretically occurs for $\mu=2$ and $\mu=6$, respectively. Here the levels $x_1=-1, x_2=0, x_3=1$ and $x_1=-2, x_2=-1, x_3=3$ form stationary but unstable states, for which $\sigma_1=0$ and therefore the coupling vanishes. In the map they would appear as vertical lines. Since we suppressed trajectories pinned on these unstable states by a small fluctuation of the initial conditions, however, they do not appear in the diagram. Indeed, the fluctuations are also necessary to see the appendix-like stability region on the lower left side of the diagrams to which the trajectories are redirected instead. We investigate this region which is accessible from the origin in subsection {\ref{sec:simjump}} with better statistics.

For $N=4$, three different three-clade distributions are possible, namely
$(2,1,1)$, $(1,2,1)$, and $(1,1,2)$. Also here, each of them is stable only
in certain intervals as can be seen from the map in Fig. \ref{fig:3splits}c) and d). $(2,1,1)$ is stable  only for $\mu$ values between 2 and 2.5 in the whole investigated
interval for $\beta$ and $\gamma$ between $-1$ and $+1$. For larger values it is mainly stable for negative coupling, whereas $(1,1,2)$ is stable for positive coupling. The distribution $(1,2,1)$ has limited stability ranges in between. Remarkably, the coupling scheme has only a minor
influence on the the distribution map.

For $N=5$, there are two 3-fold split patterns with altogether 6 distributions: $(3,1,1)$, $(1,3,1)$,
$(1,1,3)$, and $(2,2,1)$, $(2,1,2)$, $(1,2,2)$. As can be seen from Fig.~\ref{fig:3splits}e) and f) $(3,1,1)$ and $(2,2,1)$ are mainly stable for negative coupling parameters, the other distributions for positive coupling parameters. In agreement with Fig. \ref{fig:persist3splits} only the mentioned distributions survive at $\gamma=-1$ and higher values of $\mu$. For weak first order coupling there are some notable exceptions from the general rule in the form of additional stability islands for the $(2,1,1)$ and $(1,3,1)$ distributions. The lines $\beta=0$ and $\gamma=0$ are stable for all distributions since the coupling vanishes here and the system acts as five independent populations.

It should again be mentioned that metastable trajectories occur if one or more of the components $x_k$ are zero. That is not only the case for the origin as initial conditions, but also e.g. for the case that two $x_k$ are set to zero. If we do not implement fluctuations in the initial condition, we even end in four-clade states at $\mu=6$ ($\mathbf{x}=(-2,-1,0,3)$ for $N=4$ and $\mathbf{x}=(-2,-1,0,0,3)$ for $N=5$) which can be readily understood as the three-clade distribution with $\sigma_1=0$ and additional filling of zero states. They are unstable, however since any deviation in the starting conditions from zero leads to trajectories diverging from these states. Similarly, all three-clade states found below the onset of  stability contain a zero component and are truly two-clade states with a third clade that got pinned at zero. In a biological context pinning to zero ($x_k=0, \dot x_k=0$) would mean that a certain population loses its potential for change. If such a situation is to be avoided for a specific model, a drift, i.e. a non-zero value of $r$ should be included.

\subsection{Accessible Three-Clade States for $N \le 5$ by a Jump in $\mu$}\label{sec:simjump}
In the previous section we considered a wide range of starting conditions in order to show which three-clade steady states are stable attractors. Now the question should be addressed whether these states are accessible from the origin by a change in the bifurcation parameter $\mu$, either by jumping to a new environmental condition or by ramping the parameter. The origin is the only stable state for negative $\mu$, representing the single clade from which bifurcations may start. As explained above, however, we cannot start the trajectories directly at the origin since they would stay there in a metastable trajectory. Thus for each population, as in the stability maps, we again include a fluctuation by drawing a starting trait value $x_i$ from the interval $[-\epsilon/2, +\epsilon/2]$, typically $\epsilon = 0.005$.

\begin{figure}[tbp]
    \centering
    \includegraphics[width=1.0\textwidth]{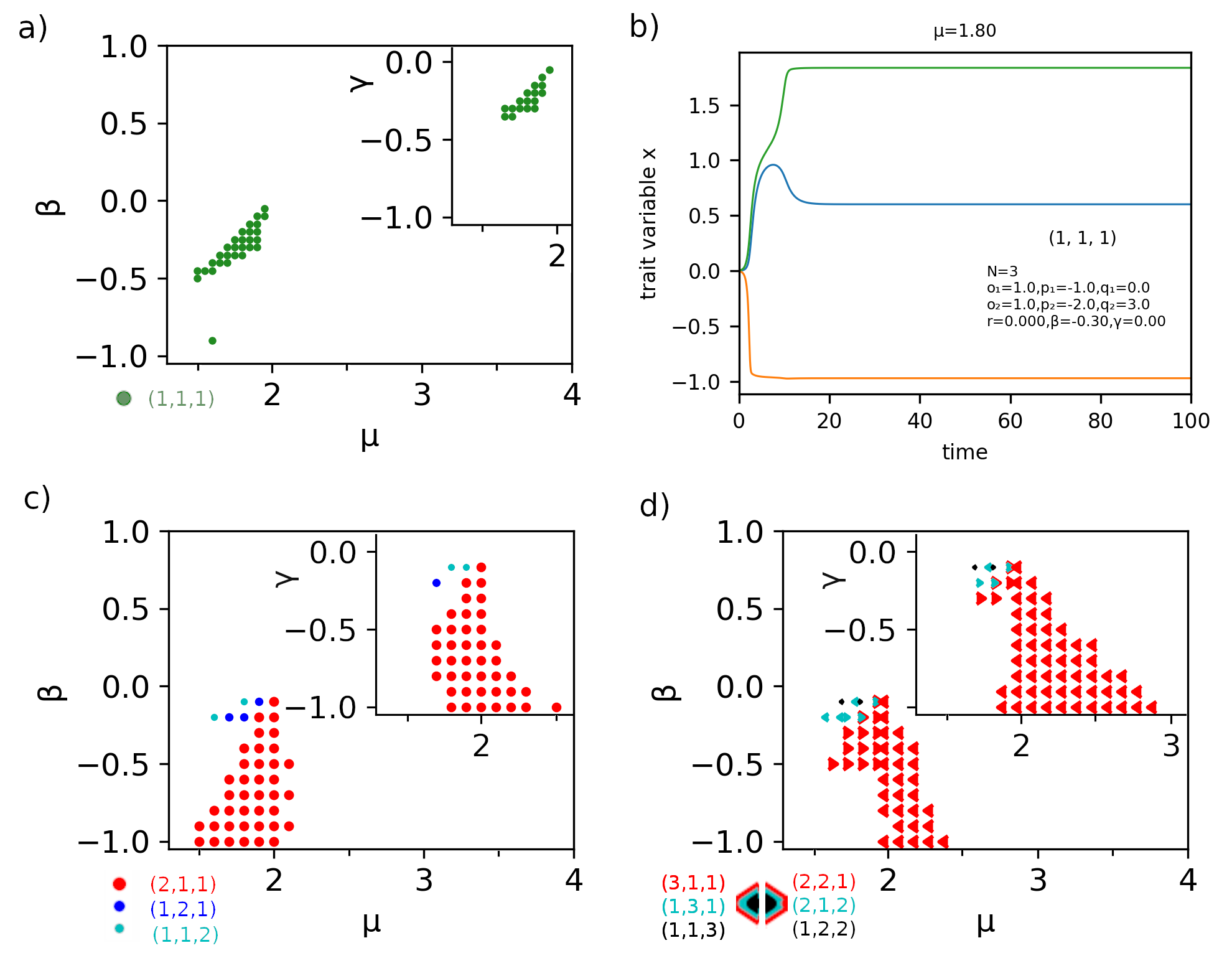}
    \caption{a) Attracting three-clade states by jumping $\mu$ from zero to a certain value (N=3). Initial values were randomly chosen from the interval $0\pm0.0025$ with 500 samples for each parameter set. Either $\beta$ or $\gamma$ (inset with same scaling) was used as coupling parameter, the other one was set to zero. b) Sample trajectory leading to a three-clade state. c) Attracting three-clade states by jumping $\mu$ from zero to a certain value for $N=4$. d) The same for $N=5$. Colour coding of the respective distributions is given as legend.}
    \label{fig:jumpmap}
\end{figure}

Again plotting $\mu-\beta$ and $\mu-\gamma$ diagrams (Fig.~\ref{fig:jumpmap}), we find a very small region in which a transition to a three-clade distribution in the case $N=3$. From this region, actually the parameters for the trajectories in Fig.~\ref{fig:moll} were chosen. It is confined to $\mu$ values below the stability onset of uncoupled populations (between 1.5 and 2) and small coupling values. The occurrence, however, is sparse with only a few events in 500 runs which indicates that the volume fraction around the origin lying within the basin of attraction of this state can be very small as Fig.~\ref{fig:moll} shows. The single point in Fig. \ref{fig:jumpmap}a at $\mu=1.6$ and $\beta=-0.9$ represents a trajectory with slowed down dynamics in which an apparent three-clade distribution is kept for a long time until it eventually undergoes a transition into a two-clade distribution. Fig. \ref{fig:jumpmap}, however, shows a typical direct trajectory to the state $\mathbf{x}=(-0.70, +0.608,+1.835)$ at $\mu=1.8, \beta=-0.3$.

For $N=4$, a similar picture arises, but here the coupling range from which a three-clade split (mainly the distribution (2,1,1)) is reached, is larger. It extends for both coupling schemes around the stability line $\mu=2$ for negative values of the coupling parameter (Fig. \ref{fig:jumpmap}c). 

For $N=5$, the accessibility region for a three-clade split (here the distribution (3,1,1)) is located more on the right side of the $\mu=2$ line, again for negative coupling parameters. (Fig. \ref{fig:jumpmap}d).

\subsection{Accessible Three-Clade States for $ N \le 5$ by Ramping $\mu$}

\begin{figure}[tbp]
    \centering
    \includegraphics[width=1.0\textwidth]{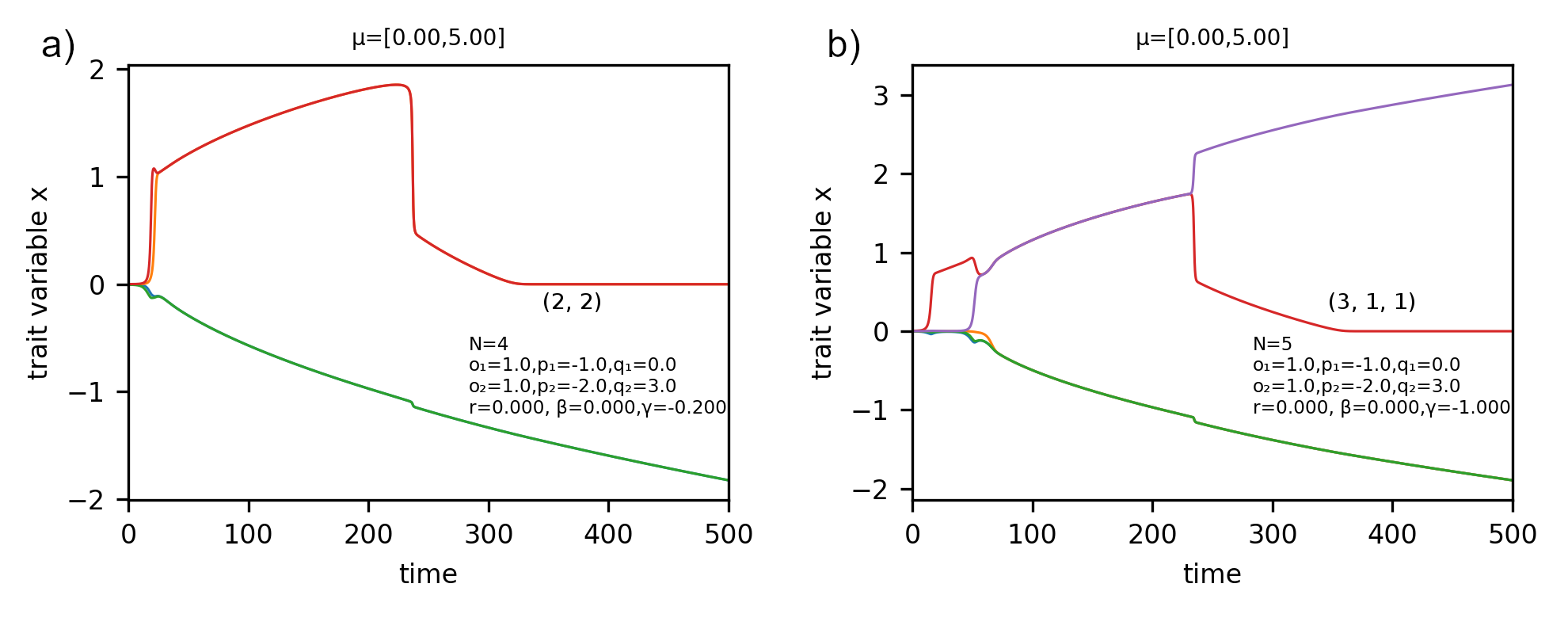}
    \caption{Selected trajectories upon ramping for a) $N=4$, b) $N=5$, respectively. In a), at the first bifurcation a (2,0,2) state is selected. At the second bifurcation the  high levels become unstable with respect to the new intermediate branch and a sudden jump to another (2,2) state occurs which, according to the levels, could be described as a (2,2,0) distribution. At around $\mu=3$ this branch crosses for one set of populations the level $x=0$ and is pinned. In contrast, in b) the distribution (3,2) undergoes a second symmetry breaking to the 3-split distribution (3,1,1). } 
    \label{fig:special45}
\end{figure}

If the environmental parameter does not change stepwise but slowly ("ramping the bifurcation parameter" \citep{gs:paper}), the situation changes. The main feature is that before the onset of the stability of a three-clade state at around $\mu=2$, the trajectory has to follow a two-clade split. So it depends whether the two-split becomes unstable and bifurcates into a three-split, which is of course not always the case. The trajectories of Fig. \ref{fig:special45} represent two examples. In \ref{fig:special45}a) for $N=4$ with a ramp of $\Delta \mu=0.01$ per time step, a two-clade state which according to the numerical values of the levels may be classified as extended distribution $(2,0,2)$ changes at  $\mu=2.4$ into another two-clade state with extended distribution $(2,2,0)$. At the transcritical intersection with stability exchange around a $\mu$ of 3 (see Fig. \ref{fig:bifmu}), the slope of the upper trajectory is changed and pinned to zero. By changing $r$ to a non-zero value, the pinning occurs at the $x=r \mu$ line. In contrast, in \ref{fig:special45}b) (for $N=5$) at around the same $\mu$ value for the state transition, a two-clade state splits into the $(3,1,1)$ distribution, followed by pinning around $\mu=3.5$. 

\begin{figure}[tbp]
    \centering
    \includegraphics[width=1.0\textwidth]{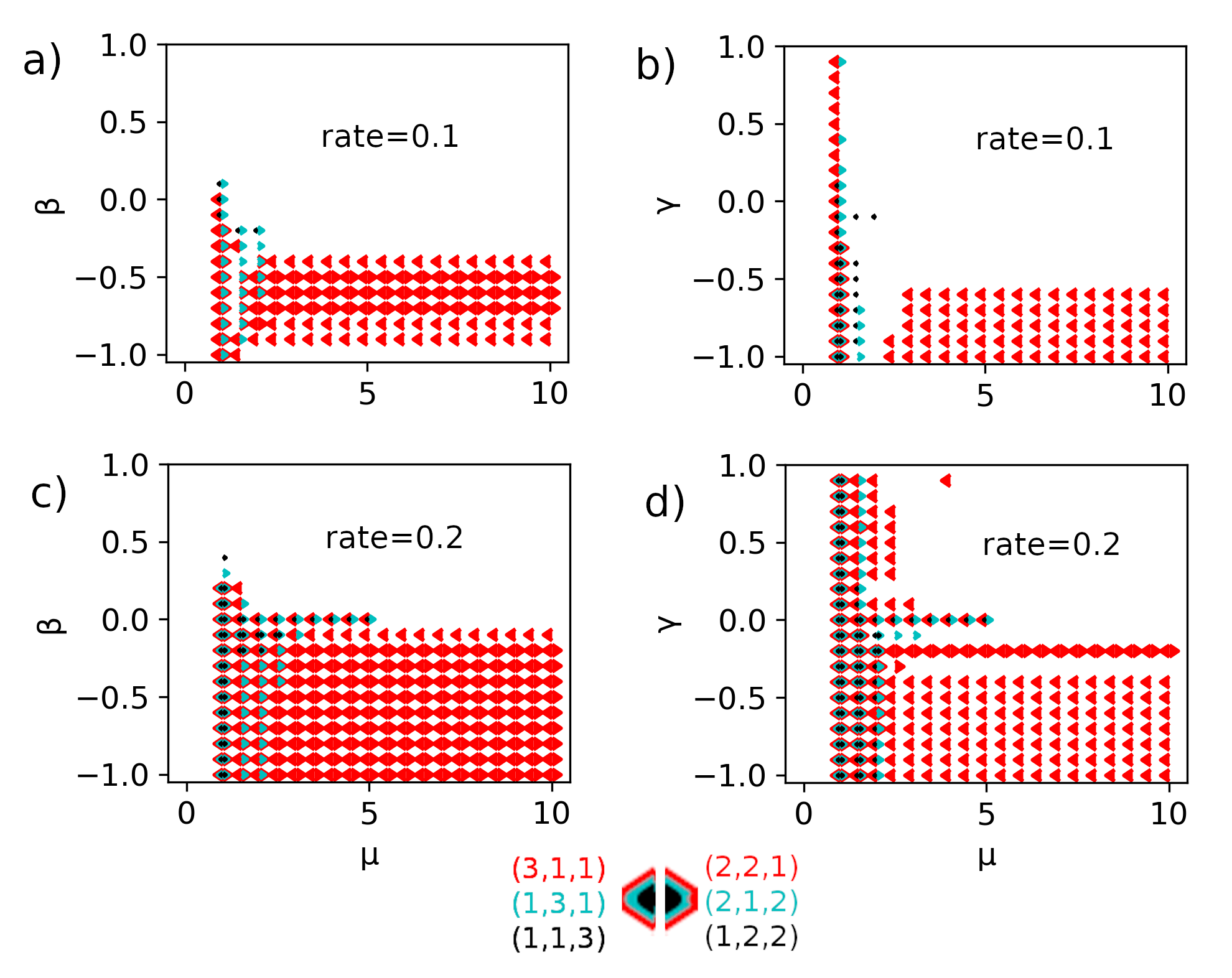}
    \caption{Occurrence of three-clade-states in ramping $\mu$ from zero to 10 at different rates (in inverse time units). Initial values were randomly chosen from the interval $0\pm0.0025$ with 1000 samples for each parameter set. Either $\beta$ (a,c) or $\gamma$ (b,d) was used as coupling parameter, the other one was set to zero.} 
    \label{fig:rampmap}
\end{figure}

As a general rule, the trajectory depends mainly on the two-clade distribution that is reached after the initial bifurcation, and here the stochastic initial conditions around the monoclade state at $\mu=0$ are decisive. After the second bifurcation, the fate of a state is more or less decided unless ongoing fluctuations are able to deter the trajectory. Small fluctuations, however, do not change the overall picture, as simulations with an adapted code revealed. Larger fluctuations that may change the transition at the bifurcation points are mimicked by higher rates (around $\Delta \mu= 0.1$) since the two-clade states are not fully equilibrated at the onset of the second bifurcation here. In general, the ramping rate is an important parameter controlling the outcome of the simulations.
Three-clade distributions may be rare, as the following example shows: For $N=5$, first order coupling with $\beta=-0.6$, an initial spread of $\pm0.0025$, and a ramp of $\Delta \mu= 0.1$ per time unit, the following distributions were obtained in 1000 runs: $(3,2)$ 87.4\%, $(4,1)$ 11.4\%, $(2,2,1)$ 0.6\%, $(3,1,1)$ 0.3\%, $(2,3)$ 0.3\%. The occurrence of the three-clade states depending on the ramp is displayed in Fig. \ref{fig:rampmap}. They occur more frequently at higher ramping rates. For lower rates ($\Delta \mu= 0.05$) almost no three-clade states are left. 

For $N=3$, independent of whether $r$ was set to $0$ or $0.001$, no 3-split was observed under ramping conditions. For $N=4$, $(2,1,1)$ distributions do occur at $\beta<-0.5$, but are rare. For example, for $r=0, \beta=-0.7$ the state $\mathbf{x}$ = (-2.691, -2.691, +0.015, +1.870) at $\mu=10$ was reached via an intermediary, not fully equilibrated (2,2) distribution. For $N=5$, the situation is much better, as the map in Fig. \ref{fig:rampmap} and Fig. \ref{fig:special45}b) as example trajectory show. Here we obtain the first trajectories that display two clearly separated bifurcations.

\subsection{Trajectory Simulations for Larger~$N$}

\begin{figure}[tbp]
    \centering
    \includegraphics[width=1.0\textwidth]{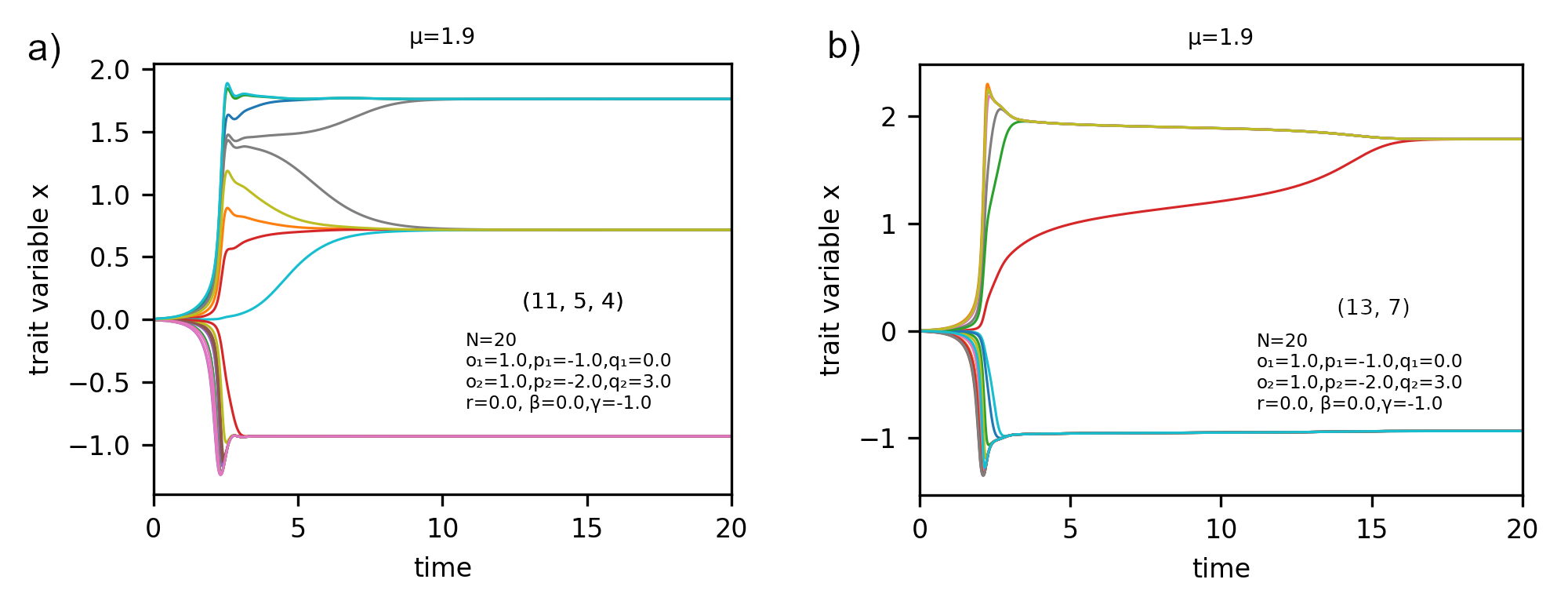}
    \caption{Two different runs for $N=20$ populations at fixed $\mu=1.9$ (jump)
      and coupling parameters $\beta=0, \gamma=-1$. Depending on the
      initial fluctuation, they result in three- or two-clade splitting,
      respectively. The distribution of the 20 populations on the observed levels is indicated in round brackets.}
    \label{fig:fixedmu}
\end{figure}

\begin{figure}[tbp]
    \centering
    \includegraphics[width=0.9\textwidth]{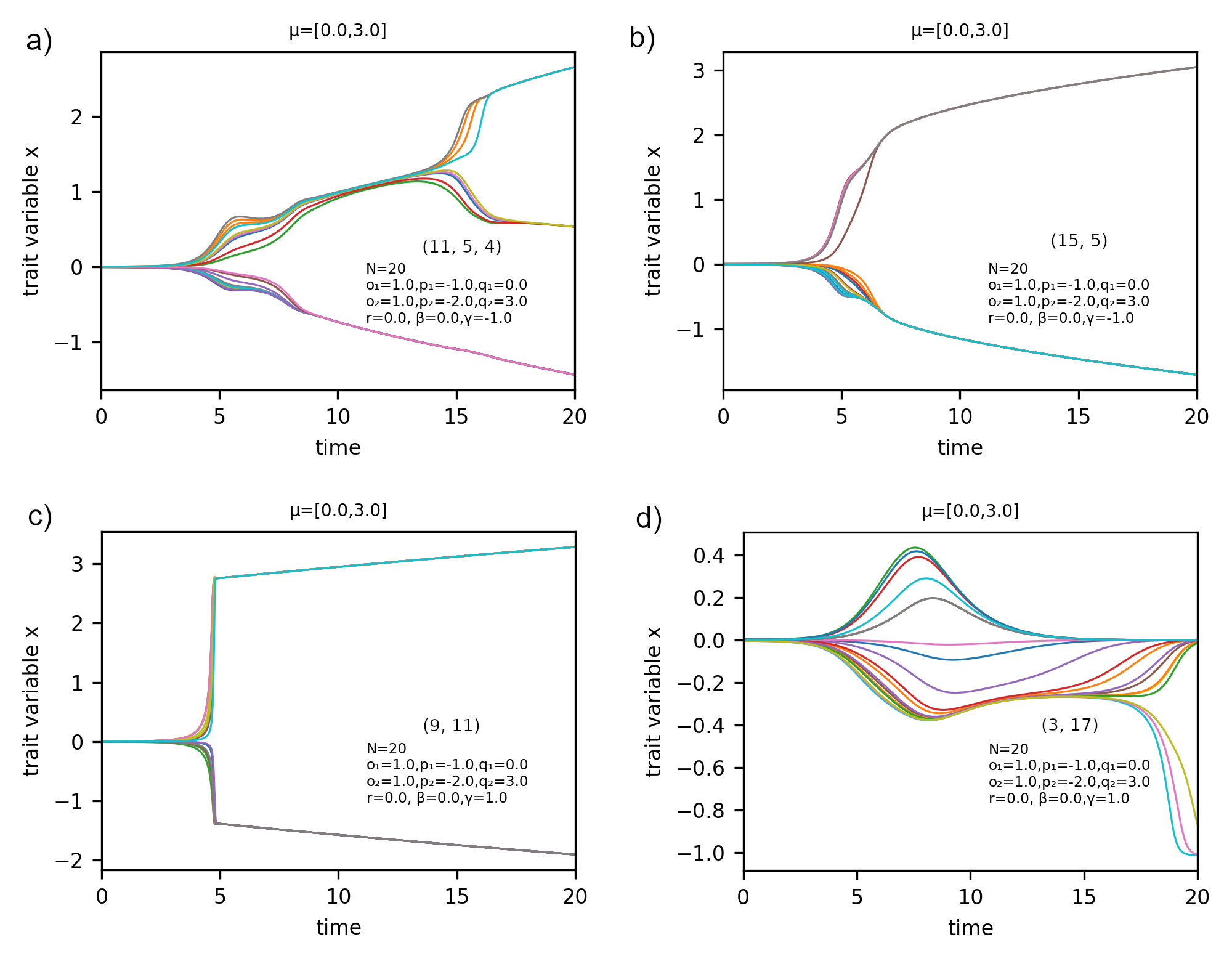}
    \caption{Two different runs each for $N=20$ populations with ramping
      $\mu$ from 0 to 3 in 20 time units either with negative or positive strong coupling (a
      and b: $\gamma=-1$; c and d: $\gamma=+1$).}
    \label{fig:ramp}
\end{figure}

The larger $N$ becomes, the more possibilities for three-clade distributions arise, and they occur more frequently. However, the occurrence still depends on the initial fluctuation. For instance, we detected for $\gamma=-1$ the transitions (4,3) $\rightarrow$ (4,2,1) for $N=7$ , (5,4) $\rightarrow$ (5,3,1) or (5,2,2) for $N=9$ and (6,4) $\rightarrow$ (6,1,3) or (6,2,2) for $N=10$, but no three-clade distribution for $N=6$ and $N=8$. Two runs with same parameters but different starting values are compared in Fig.~\ref{fig:fixedmu} for a jump and in Fig.~\ref{fig:ramp} for ramps with negative and positive second order coupling ($N=20)$. One can see the occurrence of both three-clade and two-clade distributions at the end of the trajectories. The dynamics can be very different. Sometimes the division of the populations into clades is very fast, sometimes there are "latecomers" which switch their trait after a long time. However, as Fig.~\ref{fig:ramp}a) shows, subsequent bifurcations in which first a two-clade distribution is reached, and later one of the clades splits further, is a possible scenario.

\section{Conclusions}

We have shown that symmetry breaking in dynamical systems of identical
differential equations whose right hand sides are polynomials of degree
five may lead to the desired three-clade splitting in two subsequent
events.  However, trajectories starting close to the origin do not
necessarily approach a three-clade state, even if it was possible from the
viewpoint of stability. Other stable states and separating saddle points
may form obstacles.  The probability to reach a three-clade state rises
with the number of coupled populations, since the number of three-clade
distributions rises with respect to two- or one-clade distributions.  The
model ansatz is suitable for placing bifurcation events at a certain
combination of environment $\mu$ and a trait $x$.  The assumption of $r=0$
in the growth function~$f$ of degree five has advantages in the analytical
treatment of second-order coupling, as the determination of steady states
becomes significantly easier.  But more realistic might be the assumption
that all traits vary with the environment, so that a previous value would
not become stable again. To ensure this, drifts should be included with
$r\neq0$.  However, at the present stage, such models are still far away
from quantitatively describing biological systems, but they offer important
insights in to the sympatric speciation problem. Here, we specially
investigated the case of a splitting into three clades and evaluated
methods for its analysis and finding the position of bifurcation points.

It should be mentioned that incomplete reproductive isolation as a
prototype for parapatric speciation is not only possible in space but also
in time.  This occurs in species in which the life cycle contains only a
limited timespan for gene transfer, even if for a population the temporal
distribution of reproductive activity is uniform \citep{hd:IBT}.  Examples
for organisms in which populations arise that may exhibit a pronounced
difference in their mating time include algae and other unicellular
organism that switch from a vegetative to a sexual phase such as diatoms
\citep{dedecker2019,ziebarth2023} and insects with long larval stadiums
such as annual cicadas.  Models describing such scenarios require
additional variables in order to describe cyclic behaviour and future
extensions of this work will go into this direction.

\bmhead{Acknowledgements}

Funding by the Research Training Group \emph{Biological Clocks on Multiple
  Time Scales} (GRK 2749) of the German Research Foundation (DFG)is
gratefully acknowledged.
For contributing a Python snippet we thank Arved Lieker.

\bmhead{Data Availability}

\textsc{Python} and \textsc{Maple} programs used for our analysis and for the generation of the figures and tables in this paper, as well as simulated trajectory data can be found in the publicly accessible repository DaKS (\url{https://doi.org/10.48662/daks-507}).

\appendix
\section{Linear Algebra Computations}
\label{sec:lac}

In this appendix, we detail the calculations leading to the results
presented in Section~\ref{sec:ssss}.  The main task is to compute the
determinant of the Jacobian $J$ shown in \eqref{eq:Jblock} and
\eqref{eq:JA}, as it also yields immediately the characteristic polynomial
$\chi_{J}(\lambda)$ by replacing in $\det{J}$
each $b_{k}$ by $b_{k}-\lambda$.

We subtract in each block of rows the second row from the first one, then
the third one from the second one and so on.  This brings each block of
rows in the form
\begin{equation}
  \begin{pmatrix}
    0 & \cdots & 0 & b_{k} & -b_{k} & 0 & \cdots & \cdots & \cdots & 0 \\
    0 & \cdots & \cdots & 0 & b_{k} & -b_{k} & 0 & \cdots & \cdots & 0 \\
    \vdots & & & & \ddots & \ddots & \ddots & & & \vdots \\
    a_{k} & \cdots & & \cdots & & a_{k} &  b_{k}+a_{k} & a_{k} &  \cdots & a_{k}
  \end{pmatrix}\,.
\end{equation}
Next we add in each block of columns the first column to the second one,
then the second one to the third one and so on.  Then we arrive at a matrix
\begin{equation}
  \tilde{J}=
  \begin{pmatrix}
    \tilde{J}_{1} & \tilde{A}_{12} & \cdots & \cdots & \tilde{A}_{1M} \\
    \tilde{A}_{21} & \tilde{J}_{2} & \tilde{A}_{23} & \cdots & \tilde{A}_{2M} \\
    & & \ddots & & \\
    & & & \ddots & \\
    \tilde{A}_{M1} & \tilde{A}_{M2} & \cdots & \tilde{A}_{M,M-1} & \tilde{J}_{M}
  \end{pmatrix}
\end{equation}
which has the same block structure and determinant as $J$ and the
blocks of which are of the form
\begin{equation}
  \begin{gathered}
    \tilde{J}_{k}=
    \begin{pmatrix}
      b_{k} & & & & 0 \\
      & \ddots & & & \\
      & & \ddots & & \\
      0 & & & b_{k} & \\
      a_{k} & 2a_{k} & \cdots & (N_{k}-1)a_{k} & b_{k}+N_{k}a_{k}
  \end{pmatrix}\,,\\[0.5\baselineskip]
  \tilde{A}_{k\ell}=
  \begin{pmatrix}
    0 & \cdots & \cdots & 0 \\
    \vdots & &  & \vdots \\
    0 & \cdots & \cdots & 0 \\
    a_{k} & 2a_{k} & \cdots & (N_{\ell}-1)a_{k}
  \end{pmatrix}\,.
  \end{gathered}
\end{equation}
For computing the determinant, we start now in the top left corner and
expand each row with only one non-zero entry (which is always of the
form~$b_{k}$).  This leads to the following result
\begin{equation}\label{eq:detJ}
  \det{J}=\prod_{j=1}^{M}b_{j}^{N_{j}-1}\times \det{J_{\mathrm{red}}^{(M)}}
\end{equation}
where $J_{\mathrm{red}}^{(M)}\in\RR^{M\times M}$ is the Jacobian of the
reduced system \eqref{eq:redgensys} evaluated at the corresponding
equilibrium $\bar{\mathbf{y}}$, i.\,e.\
\begin{equation}\label{eq:Jred}
  J_{\mathrm{red}}^{(M)}=
  \begin{pmatrix}
    b_{1}+N_{1}a_{1} & N_{2}a_{1} & \cdots & N_{M}a_{1} \\
    N_{1}a_{2} & b_{2}+N_{2}a_{2} & \cdots & N_{M}a_{2} \\
    & & \ddots & \\
    N_{1}a_{M} & \cdots & N_{M-1}a_{M} & b_{M}+N_{M}a_{M}
  \end{pmatrix}\,.
\end{equation}
It follows from \eqref{eq:detJ} that the characteristic polynomial of $J$
is given by
\begin{equation}
  \chi_{J}(\lambda)=
  \prod_{j=1}^{M}(b_{j}-\lambda)^{N_{j}-1}
  \chi_{J_{\mathrm{red}}}^{(M)}(\lambda)\,.
\end{equation}


\end{document}